# Evidence for Widespread Cooling in an Active Region Observed with the SDO Atmospheric Imaging Assembly

Nicholeen M. Viall and James A. Klimchuk
NASA Goddard Space Flight Center

**Abstract**

A well known behavior of EUV light curves of discrete coronal loops is that the peak intensities of cooler channels or spectral lines are reached at progressively later times than hotter channels. This time lag is understood to be the result of hot coronal loop plasma cooling through these lower respective temperatures. However, loops typically comprise only a minority of the total emission in active regions. Is this cooling pattern a common property of active region coronal plasma, or does it only occur in unique circumstances, locations, and times? The new SDO/AIA data provide a wonderful opportunity to answer this question systematically for an entire active region. We measure the time lag between pairs of SDO/AIA EUV channels using 24 hours of images of AR 11082 observed on 19 June 2010. We find that there is a time-lag signal consistent with cooling plasma, just as is usually found for loops, throughout the active region including the diffuse emission between loops for the entire 24 hour duration. The pattern persists consistently for all channel pairs and choice of window length within the 24 hour time period, giving us confidence that the plasma is cooling from temperatures of greater than 3 MK, and sometimes exceeding 7 MK, down to temperatures lower than ~ 0.8 MK. This suggests that the bulk of the emitting coronal plasma in this active region is not steady; rather, it is dynamic and constantly evolving. These measurements provide crucial constraints on any model which seeks to describe coronal heating.

## 1. Introduction

A significant outstanding issue in current solar and astrophysical research is that of the heating of the solar corona. How is the corona heated to temperatures of greater than 1 MK when the photosphere below is only 6000 K? One observational approach to addressing this important question is to focus on particular areas in the corona such as active regions (AR). Often researchers have narrowed this problem even further by analyzing discrete coronal loops within an active region. A coronal loop is an intensity enhancement relative to the neighboring plasma, observed in extreme ultraviolet (EUV) or soft X-ray wavelengths. A standard analysis approach is to isolate a single loop in an image by subtracting contributions from background and foreground emission, and then examine the loop properties such as its differential emission measure distribution (DEM), temporal behavior, spatial structure, and densities (e.g. Schmeltz et al. 2001, 2011; Warren et al. 2002, 2008; Del Zanna & Mason 2003, Winebarger et al. 2003; Winebarger & Warren 2005; Ugarte-Urra et al. 2006, 2009; Tripathi et al. 2009; Reale 2010; Aschwanden & Boerner 2011; Viall & Klimchuk 2011). Such loop studies have revealed important information about their structure and heating. It is currently accepted by most researchers that coronal loops are composed of tens to hundreds of individual sub-resolution coronal strands, where a coronal strand is a miniflux tube for which the heating and plasma properties are uniform over the cross section (e.g. Cargill & Klimchuk 1997; Warren et al. 2002, 2003; Winebarger et al. 2003; Winebarger & Warren 2005; Klimchuk 2006, 2009; Patsourakos & Klimchuk 2006), although see Mok et al. (2008) for an different explanation. This important understanding has come about through the ability of hydrodynamic models of bundles of unresolved flux tubes to reproduce EUV and X-ray emission and light curves (e.g. Warren et al.

2002; Bradshaw and Cargill 2006, 2010; Klimchuk et al. 2008). In this framework, each coronal strand is heated by an energy impulse and then cools; many such nearby strands are heated within a finite time window (a nanoflare storm), creating a coronal loop. Importantly for the analysis we will present in this paper, the light curves of these loops exhibit a common time lag pattern: the intensity of a loop in a given EUV channel (or temperature band) will increase and then decrease, with the peak intensity of the light curves occurring in sequentially cooler channels as the coronal strands cool (e.g. Warren et al. 2002; Winebarger et al. 2003; Winebarger and Warren 2005; Ugarte-Urra et al. 2006, 2009; Mulu-Moore et al. 2011; Viall & Klimchuk 2011). This is a common property of loops, even shorter loops observed in the cores of active regions (Viall & Klimchuk 2011).

These analyses have contributed greatly to our understanding of coronal heating in loops, and are an important step towards understanding the heating of active regions and the corona in its entirety. However, it is always difficult to know how representative a given loop case study is. In fact, whether a group of heated strands is even identified as a coronal loop is likely biased by the temperature band of the observation and possibly the instrument resolution. Additionally, certain line-of-sight geometries preclude even the best background subtraction methods from fully isolating the emission from a single coronal loop (Terzo & Reale 2010; Aschwanden & Boerner 2011; Viall & Klimchuk 2011).

Another limitation of these loop studies is that loops –defined observationally as the intensity enhancement relative to a more uniform background- often make up only a minority of the total AR EUV emission, with the diffuse emission in between and 'underneath' the well defined loops contributing 60-90% of the total (Del Zanna & Mason 2003; Viall & Klimchuk 2011). Importantly, this result from Viall & Klimchuk (2011) was found using the very active region we analyze here, for the same time period. Understanding this diffuse emission is therefore crucial for fully understanding the heating of active regions. It may be that the diffuse emission is simply composed of loops (defined here as physical structures), which are fading, or multiple overlapping loops, which are therefore indistinguishable. Indeed, Viall & Klimchuk (2011) presented a case study of a location of diffuse emission in an active region whose light curves exhibit cooling patterns consistent with those of the previously mentioned loop light curves, suggesting a common heating mechanism. Alternatively, it may be that the diffuse emission is produced by coronal strands that evolve entirely independently, without the collective behavior that characterizes loops. The mechanism that heats these strands could be fundamentally different from the mechanism that heats the strands of a loop. For example, the strands which comprise the diffuse corona may be heated somewhat steadily and not undergo large heating and cooling cycles as loops do (Warren et al. 2010, 2011; Winebarger et al. 2011). Therefore, in this paper we expand on these previous loop case studies and analyze all of the coronal emission in an active region, including the diffuse emission between loops, in order to identify patterns of heat release and any subsequent cooling.

Does the bulk of the coronal active region plasma exhibit cooling patterns as we have seen exhibited in countless loop studies, or is a different pattern observed for most of the active region? To address this important question, we measure intensity time lags between all possible pairs of six Atmospheric Imaging Assembly (AIA) EUV channels onboard Solar Dynamics Observatory (SDO) observed throughout an entire active region for 24 hours of data. Given the enormous number of spatial pixels involved, it is not feasible to perform a comparison by eye as has been done for loop case studies. Additionally, it is desirable to develop a more statistically rigorous way to characterize the plasma evolution. Therefore we have developed an automated procedure that involves cross correlation of intensity time series. Upon applying this procedure we find ubiquitous signatures of plasma cooling in the bulk of the active region, consistent with the expected behavior of impulsively heated plasma. The measured time lags vary systematically as a function of location within the AR, which suggests that nanoflares are more energetic in the AR core. These measurements provide crucial constraints on any model of the coronal heating mechanisms operating in this active region, whether it is nanoflares, or something else.

## 2. Methods and Analysis

We analyzed 24 hours of EUV data for NOAA active region 11082 taken with SDO/AIA (Boerner et al. 2011; Lemen et al. 2011). We plot the normalized temperature responses functions (adapted from Viall & Klimchuk 2011) of the 6 SDO/AIA EUV channels in Figure 1: we show 131 (black), 171 (cyan), 193 (orange), 211 (blue), 335 (green) and 94 (red). The 94 and 131 channels are bimodal in their temperature sensitivity, with peaks at both low and high temperatures, while 335 is broad and 171, 193 and 211 are singly peaked and narrow. These are the published response functions at the time of this writing; however there is evidence that the 94 channel in particular may have additional low temperature contributions (e.g. O'Dwyer et al. 2010; Aschwanden & Boerner 2011; Foster & Testa 2011; Testa et al. 2011). These features of the response functions (e.g. bimodality) are important for interpreting our results. As shown in Viall & Klimchuk (2011), as an impulsively heated loop strand cools, we expect the intensity to peak in the different AIA channels with different orderings that depend on the nanoflare energy. Specifically, we found that for weaker nanoflares, we expect the cool contributions to the 94 channel to dominate the light curves, and 94 should peak after 335, 211 and 193, but before 171 and 131. For strong nanoflares where the plasma is heated to much higher initial temperatures, we found that the hotter component of 94 will dominate, in which case we expect 94 to peak before all of the other channels. Regardless of the heating mechanism, if the 94 light curve rises and then falls preceding all of the others, the hot component dominates, however if the 94 light curve peaks between 193 and 171, then the cooler component dominates.

It is important to note that for impulsive heating, it is the emission from the cooling strands which dominates the light curves (Bradshaw & Klimchuk 2011). There is relatively little contribution from strands as they undergo heating because the heating phase is short lived and because the densities are relatively low. By the time chromospheric evaporation fills the strands with dense plasma making them visible, cooling is already well underway.

The active region we analyze here was located in the northern hemisphere, near disc center on 19 June 2010. In Figure 2 we show images taken at 3.5 UT in the 6 EUV AIA channels described in Figure 1. From left to right, top to bottom we display the 94, 335, 211, 193, 171 and 131 Å channels. The field of view is 450x450 pixels, where 1 pixel = 0.6”. The image intensities are all on a linear scale. This active region has a bipolar structure, was not associated with a sunspot and did not have any flaring activity above B-class for the time period that we analyzed.

We use the 30 second cadence, level 1.5 data for our analysis, for which all of the channels are coaligned and have the same plate scale. We derotate the data so that the active region emission lies in the same pixels throughout the 24 hour period. To do this, we make the simplifying assumption that the whole active region rotates at one rate, and that the rotation can be approximated by the linear projection of the angular rate in the plane of the sky. These are reasonable since the AR is near disc center and has a small latitudinal extent. Any distortion of the coronal structures due to differential rotation originates at the magnetic footpoints, and these footpoints are all rooted within 10 degrees latitude of each other. While we do not claim to have completely eliminated the effects of differential rotation, a more involved correction is not worthwhile, since the active region will experience other forms of evolution during a 24 hour period.

For the latitude of the middle of this active region, the Snodgrass empirical rotation rate (Snodgrass, 1983) is equivalent to 300 seconds per AIA pixel. With this rotation rate, we align all images with the first image on 19 June 2010; additionally, we account for fractional pixel rotation, in which emission located entirely in one pixel in the first image is split between two neighboring pixels in subsequent images. For each image we use the time change since the first image to compute the fractional number of AIA pixels that the AR has rotated. For example, the $12^{th}$ image is taken 330 seconds after the first image, and the AR has rotated 1.1 AIA pixels. We move the entire image over the integral number of pixels (for this example, 1 pixel), and we linearly interpolate between that pixel and its neighboring pixel to estimate the intensity change due to the remaining fractional pixel rotation. Linear interpolation to find the additional intensity change due to the fractional pixel rotation is crucial for time series analysis such as we perform in this paper. If the images are aligned only by integral pixel amounts, a periodic intensity change will be introduced into the time series equal to the rotation rate in seconds per pixel (for this data set, 300 seconds/AIA pixel). Note that any slight errors in our estimation of the rotation rate or differential rotation effect, will take place over time scales much longer than the transient features important for this analysis, and will not affect the results presented here.

At each pixel, in each channel, we construct a 24-hour time series from these coaligned, derotated images, which we then subdivide into smaller time windows (two consecutive 12-hour time series and twelve consecutive 2-hour time series). We linearly interpolate the full data set (the time series at all pixels, in all channels) to a common 30-s time step for this analysis. We point out that the majority of the data are already evenly sampled at exactly 30s; resampling ensures the accuracy of the time lag measurements, but changes the details of the light curves very little. For each of the 12-hr intervals and 2-hour intervals, we construct time-lag maps in the following manner. We use the IDL procedure C_CORRELATE.PRO to compute the cross correlation value of the intensity time series at one pixel in one channel with the same pixel time series observed in another channel. We compute the cross correlation values as a function of temporal offsets (up to 1 hour for the 2-hour time series, and up to 2 hours for the 12 hour time series) for both positive and negative offsets. For negative temporal offsets we use the following equation:

For positive temporal offsets we use the following equation:

where P is the cross correlation value as a function of temporal offset, L. x and y are the channel time series, and N is the number of data points. We identify the temporal offset at which the maximum cross correlation value is reached as the time lag, and construct time lag maps with the time lag measurements.

In Figures 3 and 4 we illustrate an example of this technique applied to one pixel in this AR. In Figure 3 we show the normalized light curves from 2-4 UT (7200-14400s) observed at the location indicated with the arrow in Figure 2. From bottom to top we plot the light curves observed in channels 131, 171, 193, 211, 335, and 94, each offset by 0.5 in y. There is at least one full heating and cooling cycle on strands along the line of sight visible in this time series. The intensity rises and falls in all of the channels producing local peaks, with each light curve temporally offset as a function of temperature. There are local peak intensities reached in 335 around 9000s, 211 around 10000s, 193 around 10300s UT, 94 around 10600s, and 171 and 131 around 11000s. This dynamic, time lag behavior is consistent with previous analyses of loop light curves as we described in the Introduction. Both the beginning and the end of this time series appear to capture cooling plasma as well. At the beginning of the time series, 211, 193 and 94 all decrease in intensity relative to their initial intensities, and 171 reaches peak intensity shortly after the beginning of the time series, and then also fades. This suggests that this time series began after the plasma has cooled to ~2 MK from an unknown maximum temperature at an earlier time. Similarly at the end of the time series, all of the channels exhibit intensity changes with time lags between hotter and cooler channels that are clearly associated with cooling plasma; however the time series ends before the intensities in channels 94, 171 and 131 have time to fade fully. Note that though the scale of Figure 3 highlights the distinct and systematic behavior of these local peaks in the light curves, they are at most a 40% peak-to-trough intensity change, and often only a 10% peak-to-trough intensity change.

In Figure 4, we show cross correlation values as a function of offset time between pairs of the light curves shown in Figure 3. The offset time is the amount of time the second light curve is shifted relative to the first; a positive time lag indicates that the second light curve followed the first light curve. For cross correlating the channel pairs, we generally put the hotter channel first (though 94 and 131 could be both hotter and cooler than other channels due to the bimodal nature of their response functions). The choice of which channel is first is trivial, as the opposite choice will give the same time-lag value, but of opposite sign. The peak cross correlation value for 211-193 (green) is 0.9 and is reached when the 193 light curve follows the 211 light curve by 300s (5 minutes). We record +300s at this pixel location for this channel pair in our 211-193 time lag map. The 335-211 (blue), 335-193 (orange), and 335-171 (cyan) cross correlations show similar patterns, with the peak cross correlation value reached when 211, 193 and 171 each follow the 335 light curve at 700, 1000, and 1500s time lags, respectively. This is exactly what is expected for a cooling plasma, given the respective temperature sensitivities of these channels and the approximate delays between the time series (Figure 3) that we identified by eye. 171-131 (black) reaches peak cross correlation value at no time offset between the two channels, and we record zero for the 171-131 time lag map. Lastly, we show 94-335 (red), which attains its peak cross correlation value at negative 1100s, indicating that 335 actually precedes 94, rather than follows it. This is consistent with the behavior we identified by eye in the time series. In the context of a nanoflare storm, this is consistent with a weaker heating scenario, as shown in Viall & Klimchuk (2011), in which the cooler component of the 94 channel dominates. Note that the higher noise in the 94 channel causes a lower cross correlation value, however there is still a well-defined peak in the cross correlation curve. We include the entire 2 hour time series for this cross correlation example, just as we do when we apply this technique to the entire data set, even though the beginning and end of the time series may not coincide with the beginning and end of the plasma dynamics. Though this time series seems to include more than one heating and cooling cycle, the light curves exhibit the same hotter-to-cooler time lag behavior throughout the time series, therefore the time lag signature comes through in our analysis of the cross correlation values.

## 3. Results

We perform the analysis described above to every pixel in the image set, regardless of which type of AR structure is present (e.g., fan loop, AR core, diffuse emission, long loops, short loops, hot loops, warm loops, or loop foot points). We compute the time lags between all possible pairs of the 6 EUV channels for all of the 12-hour and 2-hour datasets. We display the results in the form of time lag maps where each pixel value is the time lag associated with that particular pair of channels over that time window. For all of the maps we show a color bar, which indicates the time lag values and range. Blues, greens and blacks indicate negative time offsets where the second channel precedes the first channel; reds, oranges and yellows indicate positive time offsets, where the second channel follows the first channel; the olive green color indicates that to within the data resolution there is zero temporal offset between the two channels.

### 3.1. The 12-hr time series

We begin by presenting the results from analyzing the first 12-hour dataset. We display maps for every channel pair in Figures 5a and 5b. At the top of each panel we indicate the channel pair used. All panels are on the same color bar, except 211-193 and 171-131, which have a steeper gradient to highlight the small, but nonzero, time lag between those channels. In all 15 panels, a clear time lag signal persists. Notice that though we uniformly apply the same analysis to all of the pixels in the image, the structure of the AR is apparent in all of the maps. In the 335-211 panel (right column, middle row, Figure 5a) the active region is dominated by positive time lags, with some zero time lag, but very few negative time lag pixels. The pervasiveness of positive time lags indicates that throughout most of the active region the 335 light curve variability precedes the 211 variability; only very rarely does the 211 channel variability precede the 335. This is consistent with a scenario in which the majority of the coronal active region plasma is heated to temperatures of at least 3 MK and cools to temperatures lower than 1.6 MK. Recall that for impulsive energy release, the heating phase is much fainter than the cooling phase and does not influence the light curves.

The other three 335 maps (bottom row Figure 5a: 335-193, 335-171, and 335-131) follow a pattern consistent with that of the 335-211 map, namely positive time lags dominate the active region, largely in the same regions where they dominated the 335-211 map, with some zero time lag pixels, and very few negative time lag pixels. Comparing with the images of the AR in Figure 2 it is clear that the zero time lag locations are mostly coincident with moss, which is the transition region footpoints of hot coronal strands. The other location with many zero time lag pixels occurs in association with fan loops, though the pattern is not ubiquitous across all of the channel pairs or even throughout all of the fan-loop region. Taken together, these 4 maps indicate that throughout most of the active region, for these 12 hours, the 335 light curves precede the 211, 193, 171 and 131 light curves; plasma that reaches temperatures hotter than 3 MK is cooling down to well below 1 MK. In general, the time lags are shorter in the 335-211 map, longer in the 335-193 map, and longest in the 335-171 map. This is expected for a cooling plasma, as 211 and 193 have hotter peak sensitivities than 171.

In the top and middle row of Figure 5a we display the 94 pairs: 94-335, 94-211, 94-193, 94-171 and 94-131. As we discussed earlier, the 94 channel has significant sensitivity both at 7 MK and 1 MK. Nevertheless, we always compute time lags with 94 as the first time series in the pair. The 94-335 map mostly exhibits negative time lags in the active region, indicating that the 335 light curve variability precedes the 94 light curve variability, just as in the example pixel of Figures 3 and 4. In contrast, the core of the active region has predominately positive time lags, indicating that the 335 light curve variability follows the 94 light curve variability there. This suggests strong nanoflares in the core and weaker nanoflares outside. We see a similar pattern for the 94-211 and 94-193 pairs of positive time lags observed outside of the core of the active region, with negative time lags observed in the core. The negative time lags for the 94-211 map and the 94-193 map are smaller than 94-335, as expected from their cooler peak sensitivities. The 94-171 and 94-131 maps differ from the other three: in these maps the vast majority of the active region is dominated by positive time lags, with few negative time lag pixels. This is expected for a cooling plasma, since the 171 and 131 channels peak at cooler temperatures than the cool peak of the 94 channel. In all of the 94 maps the time lag values at the loop footpoints, or the moss, generally are consistent with zero time lag.

In Figure 5b we show the 6 remaining channel pairs: 211-193, 211-171, 211-131, 193-171, 193-131 and 171-131. The 211-193 map exhibits almost exclusively a positive or zero time-lag, with 193 following 211, and there is a clear tendency for the time lag to increase with distance from the core. Importantly, there is virtually no occurrence of negative time lags (193 preceding 211), even very small ones. The 211-171, 211-131, 193-171 and 193-131 maps show similar patterns to those present in the 335 maps, namely the cooler channels follow the hotter channels (indicated with a preponderance of positive time lags) for the majority of the active region. In the moss regions we find time lags consistent with zero in these maps too. Lastly, in the 171-131 time-lag map (also with a compressed color table) almost the entire active region is correlated near zero time lag, as we found in the single pixel example presented in Figure 4. There are some regions with slight positive time lag, and there are some regions with slight negative time lag, but zero time lag pixels dominate much more than in any of the other maps.

We repeat this analysis on the second 12 hours on 19 June 2010 and find qualitatively similar results in the maps: positive time lags dominating all but the first three 94 maps; progressively larger time lags as the peak temperature sensitivities of the channel pairs are further apart; a 94 positive-to-negative inversion signature in the core of the active region; and zero time lag at the moss footpoints.

### 3.2. The 2-hr time series

We repeat the analysis, dividing the data into twelve consecutive 2-hour windows. This tests the degree to which single, short-lived intensity variations, surrounded by otherwise steady emission, dominate the 12 hour results. Do the 2-hour results display qualitatively the same results? We display time lag maps in Figure 6 computed using the first 2 hours of data in the same format as the 12-hour maps. We calculate cross correlation values for up to an hour temporal offset in the positive and negative direction, rather than a full 2 hours, for these shorter time windows. As the temporal offset increases, the number of data points used to compute the cross correlation decreases. Any temporal offsets greater than half the data window (in this case, greater than 1 hour), are computed using the variation of less than half of the time series, and therefore are much less meaningful.

Qualitatively these results are the same as the 12 hour results, though there is more noise in these results. The 94-335, 94-211 and 94-193 channel pairs exhibit negative time lags in most of the active region, with positive time lags in the core, while the other 6 panels in Figure 6a are dominated by positive time lags throughout the active region. In 6b we see that positive time lags persist for the majority of the active region in the 211-171, 211-131, 193-171 and 193-131 maps. Again, the 211-193 and 171-131 maps (Figure 6b) are displayed using a steeper color gradient to highlight the small but non-zero time lags. In 211-193 there is a larger occurrence of negative time lags than in the 12-hr maps, but the map is still dominated by positive time lags. The 171-131 map has fewer olive green pixels than the 12-hour version, but as in the 12-hour version there is a mix of positive and negative time lags. All of the maps have zero time-lag correlations near the magnetic footpoints, as in the 12-hr maps.

These patterns persist for all 12 of the 2-hour maps in this 24 hour period. To illustrate this, in Figure 7 we show the 2-hour 335-211 maps for all twelve intervals in our dataset. All 12 maps are consistent with a pattern of ubiquitous cooling for the duration of the 24 hour period. The details change across the different panels, and the exact time lag value at a given pixel may vary from map to map. This is due to different flux tubes, or groups of flux tubes, with different properties dominating the emission in a particular pixel and 2-hr window. However, a basic pattern of positive time lags persists across the entire 24 hour period. In general, pixels for which positive time lags are measured have positive time lags in all 12 of the maps and regions of zero time-lag almost exclusively occur near the moss.

There is a trade-off between window length and noise suppression. Comparing Figure 5 with Figures 6 and 7 it is clear that the 2-hour results contain more noise, simply due to having fewer total variability cycles per time series to cross correlate. This is also evident in the cross correlation values themselves, which are lower for the 2-hr time series (see Appendix). Additionally, the 2 hour windows are more susceptible to 'edge effects', where the time series captures only part of the plasma variability. This occurs because the start time of the windows are arbitrary (every two hours, on the hour), and may have nothing to do with the physical beginning of any dynamics. For example, in the nanoflare storm scenario, if a pixel time series contains only the very end of one nanoflare storm, followed by only the very beginning of another, spurious time lags may result. In such a pixel the brightening and fading of 211 may even 'precede' that of 335, simply because the 335 peak from the first storm occurred before the time window begins and is missed, and likewise the 211 peak associated with the second storm occurred after the time window ends and is missed. In pixels with spurious edge effects, the measured time lag is not equivalent to a physical 'cooling time' of the plasma. As the time window increases, the likelihood of capturing many full cycles of plasma variability increases, so these spurious edge effects as well as noise effects will be less and less significant.

The 12 hour maps have reduced effects from noise and reduced edge effects; however there may be a question of persistence of the cooling. Perhaps the 12-hour maps are dominated by the time lag cooling signal simply because at some brief time in those 12 hours there was a feature which cooled through those pixels. With our 2-hour maps we can immediately rule out this scenario. If this were the case, then we would expect only the 2-hour window that contained the transient feature to exhibit the time lag cooling signature. The 2-hour window results (Figure 7) demonstrate that it is generally not a single, brief, transient event (such as one very bright loop) along the line of sight which dominates the full 12 hours. We see qualitatively similar behavior for all of the sub-windows for the entire 24 hour time period and throughout the active region demonstrating that the cooling behavior is continual.

We stress that the cooling is widespread within the active region and, with the exception of moss and fan structures, is observed in most pixels. This includes pixels that are intersected by discernable loops as well as those that are not. The diffuse component of the corona is observed in pixels without loops and it is also seen as the "background" emission that dominates pixels with loops. As discussed in the Introduction, the diffuse component could represent several overlapping and therefore indistinguishable loops, or it could result from many strands heated completely independently without any spatial coherence. This is an important and open question.

The time lag maps (Figures 5-7) seem to follow the general structure of the active region as defined by the magnetic field. This is expected for both discernable loops and the diffuse component, since plasma is organized along the magnetic field in both cases. We note that the width of the loop-like features (features that have the same or very similar time lag results across their width) in the time lag maps ranges from the instrumental resolution up to ~25 pixels, or 1.2-15". The large end of this range is generally much greater than the width of observed loops; for example Watko & Klimchuk (2000) and Aschwanden & Nightingale (2005) measured TRACE EUV loops widths to be typically ~ 3.4" and 1.5-4.2", respectively, and Klimchuk (2000) found even the hotter, lower resolution SXT loops to be only 12". Though some of the narrow features no doubt correspond to individual discernable loops, the magnetically aligned features in the time lag maps cannot simply be interpreted as individual loops.

## 4. Discussion

Our main interpretation of these measured time-lags is that the plasma must be cooling throughout most of the active region for the full duration of the 24 hours. Furthermore, because positive time lags are widespread for the 335-211, 335-193 and 335-171 time lag maps with the longest time lags observed in the 335-171 maps, it is clear that at least some plasma must be heated to temperatures of greater than 3 MK, and must be cooling down to temperatures less than 0.8 MK before undergoing any further heating. If there were positive time lags only present in the 335-211 and 193-171 maps, but not in the 335-171 maps, that could indicate two separate populations of plasma: one population which cools down from temperatures greater than 3 MK, but which heats back up before it cools below 1.6 MK, and a different population which never gets hot enough to emit strongly above 1.6 MK before cooling down.

These observations of the entire active region are consistent with the known cooling behavior of coronal loops as we discussed in the Introduction, as well as consistent with our nanoflare storm models coupled with the AIA response functions discussed here and shown in Viall & Klimchuk (2011). In general, the value of the time lags calculated here are also consistent with predictions of AIA light curves from the nanoflare storm models and case studies of Viall & Klimchuk (2011). For example, they showed that the time lag between 335 and 171 could be near 1100s for short core loops and 2500s for longer loops. This is a typical range of expected time lags, though the exact value of the time lags associated with nanoflare heated plasma could be even more or less than these values, depending on the flux tube length, the energy of the nanoflare, and the initial density of the flux tube. These observations presented here are also consistent with a recent study of the distributions of soft X-ray intensity fluctuations (Terzo et al. 2011) which suggests widespread cooling. It should be noted that a given AIA pixel likely contains emission from many hundreds to thousands of flux tubes, whose behavior may be entirely independent of one another. Those flux tubes which are emitting in these AIA channels and are dynamic will all contribute to the single measured time lag for that pixel and that time window analyzed. Therefore the measured time lag could be a composite or average cooling time, rather than the cooling time of a single monolithic structure. Additionally, the 12-hr window pixels potentially have individual flux tubes undergoing multiple cycles of heating and cooling, and each cycle may have different heating properties. Finally, the peak cross correlation in some pixels could identify longer term cooling trends, rather than the cooling of individual flux tubes. We examine this possibility in our companion study.

Even the more complex behavior exhibited by the 94 pairs is easily understood in the context of the nanoflare models, the bimodal nature of the 94 temperature response, and the other, more straightforward maps. As we discuss in Viall & Klimchuk (2011), the only way for 335, 211 and 193 to precede 94 (the negative time lags in those maps) but for 171 and 131 to follow 94 (positive time lags) is if the cooler, 1 MK component of 94 is dominating the light curves of those pixels. Likewise, the only way for 335, 211, 193 and 171 to all follow 94 is if the hot component of 94 dominates the light curves of those pixels. All of this suggests that the plasma in the longer flux tubes, away from the core of the active region have 94 light curves dominated by the 1 MK plasma, while the shorter, core flux tube light curves tend to be dominated more by the hotter component of 94, as found in Reale et al. (2011). Recalling the channel order predictions of Viall & Klimchuk (2011), this implies stronger nanoflares in the core of the active region and weaker nanoflares outside, consistent with the interpretation of Reale et al. (2011). This is not surprising, since nanoflares are likely to be magnetically driven, and the magnetic field strength decreases with distance from the center of the active region. In contrast to the 94 channel, we only seem to find evidence for the cool component of 131, suggesting that the coronal plasma in this active region does not strongly emit at temperatures greater than ~ 10 MK. Such ultra-hot plasma either has a very small emission measure or is strongly influenced by nonequilibrium ionization effects (e.g., Reale & Orlando 2008; Bradshaw & Klimchuk 2011).

The nanoflare model naturally explains the observations presented in this paper, including the time lag inversion in the 94 maps. What about other heating scenarios - can they too reproduce all of the aspects of the time-lag maps? For example, how could steady heating in which the plasma is constantly maintained at a single temperature and never has a chance to cool produce such time lag signatures? In this case the emission is by definition time stationary, and therefore the variability and resulting cross correlation will be exclusively due to noise. Noise will produce approximately equal amounts of positive and negative time lags, those time lag values will be random, and zero time lag will occur no more often than any other particular time lag value. Furthermore, our maps exhibit clear spatial organization in that the basic AR structure is visible; noise is not expected to have any particular spatial organization. Lastly, time lags resulting from noise would not produce the observed increase in time lags as a function of channel separation. We can be confident that for this active region, for these 24 hours, the majority of the emission cannot be due to truly steady heating.

While we can be confident that the majority of the emission is due to cooling plasma, and that the amount of steady emission must be small, we cannot determine exactly what fraction of the plasma along a given line of sight is dynamic instead of steady without the use of models. Some of the emission in this active region may be truly steady. In our companion paper we use the time lag and cross correlation value measurements presented in this paper in combination with models of line-of-sight integrated emission from a combination of nanoflare heating, steady emission, and noise to evaluate the fraction of emission in a given pixel that could be due to steady heating.

There are heating scenarios other than impulsive nanoflare heating which may also be able to reproduce some of the time lag behavior. For example, truly steady heating can produce time-variable emission under thermal non-equilibrium conditions (e.g., Klimchuk, Karpen & Antiochos 2010; Lionello et al. 2011), though generally on small time scales. Another possibility, type II spicules, which inject heated chromospheric material into the corona (De Pontieu et al. 2011), may also contribute hot plasma that may produce a cooling signature such as those we observed here. A third possibility is a variation on the basic nanoflare model, in which the heating is quasi-steady in the sense that the nanoflares repeat frequently enough on each strand so that the plasma only cools partially before being reheated. This would not, however, explain the observed time lags between channels that are widely separated in temperature. In our companion paper we model a range of nanoflare frequencies for direct comparison with these time lag results. In that paper we also consider the possibility of nanoflares initiating randomly on physically separate coronal strands along a single line of sight, without the coherence expected for a bundle of strands comprising a loop. Regardless of the outcome, the results we present here are important constraints on all models of heating of the active region coronal plasma.

Finally, there are significant moss/footpoint areas where the time lags are consistent with zero in most if not all of the channel pairs. This result is expected for the transition region emission from impulsively heated strands. For individual strands, the light curves in a particular channel from cooling coronal plasma are narrow, and the peaks have a clear separation in the different channels. This is not true for the emission coming from the transition region. Transition region light curves are much broader (the emission persists for a longer time in each channel than do the corresponding coronal light curves), and because the intensity tends to scale with pressure at all temperatures throughout the transition region, the channel intensities all peak at roughly the same time (when the pressure peaks). This results in much greater overlap and temporal correspondence of all of the light curves. Consequently, we expect time lag signatures to be much less pronounced in the footpoints of impulsively heated strands than in pure coronal observations, just as we observe in our time lag maps. The other location in the AR that exhibited time lags consistent with zero was the fan loop area, though it was not as pronounced as the moss area. A possible cause is transverse waves carried on those strands. Transverse waves will tend to raise the cross correlation value at zero temporal offset wherever they are present and the emission from the strands which carry them is significant in both channels. This is simply a consequence of the wave bringing the strands (and their emission) in and out of the field of view of the pixel. Finally, we also observe a significant amount of zero time lag pixels in the 171-131 maps. This could indicate that though the plasma is cooling, it is not cooling much below the 171 peak temperature of 0.8 MK. Alternatively, it may be that the plasma cools so rapidly from 0.8 to 0.5 MK (for example, if enthalpy cooling dominates) that we cannot detect a time lag with the 30s resolution data that we use. A third possibility is that the contributions to the 131 channel near 0.7 MK are much more than the current response functions indicate (e.g., Testa et al. 2011).

## 5. Conclusion

We measure time lags between the AIA coronal channels across the entire active region, including distinguishable loops and the emission between loops. These measurements provide crucial constraints on any model which seeks to describe the heating and subsequent cooling of coronal plasma. Our main findings are:

1) We observe cooling plasma throughout the active region, for the entire 24 hour duration analyzed, independent of time series window size. The cooling plasma is observed both in discernable ‘loops’ as well as in the diffuse emission of the active region. Though we apply the same analysis to all of the pixels in the image without regard to AR feature, the basic AR shape is apparent in all of the maps.

2) Time lags are observed between almost all channel pairs, including those with large temperature separation (e.g., 335-171), and time lags increase between channels with wider temperature separations. This indicates that coronal strands cool fully from temperatures greater than at least 3 MK, and sometimes as high as 7 MK, down to temperatures lower than ~ 0.8 MK.

3) The 94 results exhibit systematic behavior as a function of location in the active region. We find that the hot, 7 MK component of the 94 channel dominates the light curves of the core of the active region, while the cool, 1 MK component dominates outside the core, which suggests that nanoflare energies decrease away from the active region core (Viall & Klimchuk 2011).

4) Time lags are small for lines of sight that include moss, the transition region footpoints of hot (>3 MK) coronal strands, which is also consistent with impulsive nanoflare heating.

A significant amount of the emitting plasma is not steady; rather, it is dynamic and evolving. We discuss these results in the context of impulsive nanoflare heat release where the plasma cools fully before it is subsequently reheated, as shown in Viall & Klimchuk (2011). Our results are not consistent with the majority of the emission being caused by truly steady heating which produces steady emission. In a companion paper we examine these results in the context of effects of line of sight integration where physically separate strands along the same line of sight undergo out-of-phase impulsive heating. Finally, in this paper we have only analyzed a single active region, and there is no guarantee that other active regions, or even this active region later in its evolution, behave the same way. We are currently investigating other active regions in various phases of their evolution to address these questions.

## 6. Appendix

In this section we show the time lag results shown in Figures 5 and 7, reproduced in histogram form. We also show histograms of the cross correlation values associated with these results, as well as an example of a cross correlation value map. In Figure A.1 we show histograms of the time lags recorded in map-form in Figure 5, and in Figure A.2 we show histograms of the time lags recorded in Figure 7. For every histogram we list the total number of positive, negative, and zero time lag pixels in the panel. The x-axis is the time lag in seconds, the bins are always 30 second bins, and the y-axis is the log number of pixels (note that the y scale is not the same for each panel). In the upper left corner of each panel we list the channel pair.

Beginning with Figure A.1, all four panels with 335 as the first channel exhibit an asymmetry in the histogram, where positive time lags greatly outnumber negative time lags nearly two to one. This indicates that variability in 335 precedes variability in 211, 193, 171 and 131, entirely consistent with the maps shown in Figure 5. There are a greater number of short positive time lags in 335-211 and 335-193 than there are in 335-171 and 335-131, consistent with the fact that 171 and 131 are separated further from 335 in peak temperature than 211 and 193 are. The 94 pairs generally exhibit a broader peak around zero time lag, but still exhibit asymmetric time lags. For 94-335, 94-211, and 94-193, there are more negative time lags. The asymmetry switches in 94-171 and 94-131, with these pairs exhibiting more positive time lags. Recall that this is due to the cooler component of the 94 channel, which dominates the majority of the active region, with the exception of the core emission. The core region hot emission produces positive time lags, which for 94-335, 94-211 and 94-193 causes the negative time lag asymmetry to be less pronounced at short time lags. The 211-193, 211-171, 211-131, 193-171, and 193-131 pairs all exhibit clear, large asymmetries with the number of positive time lag pixels greatly exceeding the number of negative time lag pixels by more than a factor of two or three, and in the case of 211-193, by a factor of four. The 171-131 histogram has the highest number of zero time lag pixels of any of the other pairs, and there are slightly more negative time lag pixels than positive. Lastly we note that there is a large peak in all of the histograms at zero time lag; to be expected based on the large number of zero time lag pixels present near the moss. All of the positive time lag pixels are spread out between the 250 possible time lag bins in this histogram (likewise, all of the negative time lag pixels are spread out between 250 negative time lag bins), while all time lags exactly equal to zero, plus all of those with time lags less than the resolution of the data are in the single zero time lag bin. All of these histograms are entirely consistent with the time lag maps discussed in the main text. This is to be expected, as they consist of the same information presented in the time lag maps, however these are a complementary way to visualize the results and to see the clear persistent pattern of cooling.

In Figure A.2 we show histograms of all 12 of the 2-hour window 335-211 results, complementary to Figure 7. The histograms are in the same format as Figure A.1, however note that there are fewer time lag bins, due to the fact that we only compute the cross correlation value for +/- one hour time lag. The y-axis is the same for all 12 histograms. Though the exact number of pixels in a particular positive time lag value bin varies from interval to interval, all 12 of the histograms are extremely similar. There is a clear asymmetry where the positive time lag dominates the negative time lag in all 12 intervals, confirming that this effect is not the result of a few transient events dominating only particular time intervals.

One measure of the uncertainty in our results is its stability and persistence, as we demonstrated in our 2-hr results shown in Figures 7 and A.2. Statistically speaking, the result is the same for all 12 of the 2-hr sub windows: a large number of pixels contain cooling plasma as the dominant source of their variability, and the spatial pattern of the time lag measurements (e.g. zero time lag in the moss and mostly positive time lags inside the AR) persists through all 12 maps. A different measure of the uncertainty in these results is the cross correlation values of the time lags.

We present histograms of the cross correlation values that correspond to the time lag results presented in Figure 5 in Figure A.3, and those that correspond to Figure 7 in Figure A.4. The bin size is 0.01. The first striking feature to note is that there are very few instances of zero or negative cross correlation value in any of the histograms. It is almost never the case that the intensity time series of two channels are anticorrelated. The other striking feature is that the cross correlation value is mostly dependent on the count rate of the channel: time lag measurements made with those channels with high count rates (e.g. 211 and 193) have very high cross correlation values, those made with channels that have low count rates (e.g. 94 and 131) have lower cross correlation values. Still, even the channel pairs with the lowest correlations like 94-131 pair have a large number of pixels with cross correlation values greater than 0.3, and channel pairs such as 211-193 are highly correlated, with almost no values below 0.2. A.4 shows the cross correlation values for the twelve 2-hour window 335-211 results. As expected, the result changes little from window to window. Also, as discussed in section 4, these shorter windows produce lower cross correlation values and are in general noisier than longer windows.

Lastly, we display in the top left panel of Figure A.5 a map of the cross correlation values associated with the 94-335, 0-12 UT window, time lag map shown in Figure 5 and the histogram shown in A.3. Though the histogram shows that this channel pair results in the most pixels whose cross correlation values are low (near 0.2), the map shows that the majority of these low cross correlation pixels occur in the corners of the image, outside of the main part of the active region, due to very low count rates in these areas in both 94 and 335. Though the overall occurrence of high cross correlation values is less frequent in the 94-334 map, high cross correlation values occur frequently in the main area of the active region, where the count rates are higher. The 94-335 time lag map (Figure 5a) exhibits little coherent spatial behavior in the corners of the image, where the cross correlation values are low. In contrast, the 94-335 time lag map exhibits much coherent spatial behavior in the main body of the active region where the cross correlation values are high. We show this explicitly in the top right panel of Figure A.5. We display the time lag map from Figure 5 with all pixels with cross correlation values less than or equal to 0.2 blacked out. The corresponding histogram of time lag values only at pixels where the cross correlation value is greater than 0.2 is shown in the lower panel. This cross correlation value threshold discards two thirds of the pixels, keeping only the highest third (106150 pixels are discarded, having too low a cross correlation value). Still, the basic time lag inversion in the core is observed, as well as the coherent spatial structure in the core, and the asymmetry in the histogram of the time lags. This gives confidence that these results are robust, as the other channel pairs generally exhibit much higher cross correlation values, and the conclusions of this paper are primarily based on this main area of the active region where the cross correlation values are high in all channel pairs.

Acknowledgements
The research of NMV was supported by an appointment to the NASA Postdoctoral Program at the Goddard Space Flight Center, administered by Oak Ridge Associated Universities through a contract with NASA. The research of JAK was supported by the NASA Supporting Research and Technology program. The data are courtesy of NASA/SDO and the AIA science team. We thank the referee for their helpful comments.

References

Aschwanden, M. J., & Boerner, P. 2011, ApJ, 732, 81.
Aschwanden, M. J., & Nightingale, R. W. 2005, ApJ, 633, 499.
Boerner, P., Edwards, C., Lemen, J., et al. 2011, Sol. Phys.
Bradshaw, S. J., & Cargill, P. J., 2006, A&A, 458, 987.
Bradshaw, S. J., & Cargill, P. J., 2010, ApJ, 717, 163.
Bradshaw, S. J., & Klimchuk, J. A., 2011, ApJ S, 194, 26 doi:10.1088/0067-0049/194/2/26.

Cargill, P. J., & Klimchuk, J. A. 1997, ApJ, 478, 799.

Del Zanna, G., & Mason, H. E. 2003, A&A, 406, 1089.
De Pontieu, B., et al. 2011, Science, 331, 55.
Foster, A. & Testa, P., 2011, ApJ Letters, 740:L52.
Klimchuk, J. A. 2000, Sol. Phys., 193, 53.
Klimchuk, J. A. 2006, Sol. Phys., 234, 41.
Klimchuk, J. A. 2009, in Second Hinode Science Meeting: Beyond Discovery-Toward Understanding (ASP Conf. Ser. 415), ed. B. Lites et al. (San Francisco, CA: ASP), 221.
Klimchuk, J. A, Karpen, J. T, & Antiochos, S. K. 2010, ApJ, 714, 2.
Klimchuk, J. A., Patsourakos, S., & Cargill, P. J. 2008, ApJ, 682, 1351.
Lemen, J., Title, A. M., Akin, D. J., et al. 2011, Sol. Phys.
Lionello, R., Winebarger, A., Mok, Y., Linker, J. & Mikic, Z. 2012, ApJ, Submitted.
Mok, Y., Mikic, Z., Lionello, R., & Linker, J. A., 2008, ApJ, 679, L161.
O'Dwyer, B.; Del Zanna, G.; Mason, H. E.; Weber, M. A.; Tripathi, D. 2010, A&A, 521, A21.

Mulu-Moore, Fana M.; Winebarger, Amy R.; Warren, Harry P.; Aschwanden, Markus J. 2011, ApJ, 733, 59.
Patsourakos, S. & Klimchuk, J. A., 2006, ApJ, 647, 1452.
Reale, F., 2010 Living Reviews in Solar Physics, 7, 5.
Reale, F. & Orlando, S., 2008, ApJ, 684,715.
Reale, F., Guarrasi, M., Testa, P., DeLuca, E. E., Peres, G., Golub, L. 2011, ApJL, 736, L16.
Schmelz, J. T., Jenkins, B. S., Worley, B. T., Anderson, D. J., Pathak, S., & Kimble, J. A. 2011, ApJ, 731, 49.
Schmelz, J. T., Scopes, R. T., Cirtain, J. W., Winter, H. D., & Allen, J. D. 2001, ApJ, 556, 896.
Snodgrass, H. B. 1983, ApJ 270, 288.
Terzo, S. & Reale, F. 2010, A&A, 515, A7.
Terzo, S., Reale, F., Miceli, M., Klimchuk, J. A., Kano, R., and Tsuneta, S. 2011, ApJ, 736, 2.
Testa, P., Drake, J. J., Landi, E. 2012 ApJ accepted.
Tripathi, D., Mason, H. E., Dwivedi, B. N., DelZanna, G., & Young, P. R. 2009, ApJ, 694, 1256.
Ugarte-Urra, I., Warren, H.P. and Brooks, D.H., 2009, ApJ, 695, 642–651.
Ugarte-Urra, I., Winebarger, A. R., & Warren, H. P. 2006, ApJ, 643, 1245.
Viall, N. M., & Klimchuk, J. A. 2011, ApJ, 738, 24.
Warren, H. P., Brooks, D. H., and Winebarger, A. R., 2011, ApJ, 734, 2.
Warren, H. P., Ugarte-Urra, I., Doschek, G. A., Brooks, D. H.,&Williams, D. R. 2008, ApJ, 686, L131.
Warren, H.P., Winebarger, A. R., and Brooks, D. H., 2010, ApJ, 711, 228.
Warren, H. P, Winebarger AR, and Hamilton, 2002 PS, ApJ, 579, L41.
Warren, H.P., Winebarger, A.R., and Mariska, J.T., 2003, ApJ, 593, 1174–1186.
Watko, J. A., & Klimchuk, J. A. 2000, Sol. Phys., 193, 77.
Winebarger, A. R., & Warren, H. P. 2005, ApJ, 626, 543.
Winebarger, A. R., Schmelz, J. T., Warren, H. P., Hulburt, E. O. , Saar, S. H., Kashyap, V. L., 2011, 740, 1.
Winebarger, A.R., Warren, H.P. and Seaton, D.B., 2003, ApJ, 593, 1164–1173.

**Figures Captions**

**Figure 1.** Normalized temperature response functions (adapted from Viall & Klimchuk 2011) for 6 SDO/AIA EUV channels: 131 (black), 171 (cyan), 193 (orange), 211 (blue), 335 (green) and 94 (red).

**Figure 2**. NOAA AR 11082 in six SDO/AIA channels at 3.5 UT on 19 June 2010. Upper left corner indicates peak temperature sensitivity of that channel. From left to right, top to bottom the images show channels 94, 335, 211, 193, 171 and 131. The images are displayed on a linear scale. The corresponding grey scale ranges in counts/s are 0-32, 0-207, 12-1746, 0-3637, 22-2492, and 0-113, respectively. 1 pixel = 0.6” and an arrow indicates pixel used for Figure 3.

**Figure 3**. Normalized light curves over two hours of location indicated with arrow in Figure 2 images. From bottom to top we plot channels 131, 171, 193, 211, 335, and 94, each offset by 0.5 in y.

**Figure 4.** Cross correlation values as a function of offset time between pairs of light curves shown in Figure 3. 211-193 (green), 335-211 (blue); 335-193 (orange); 335-171 (cyan); 171-131(black) 94-335(red). Dots indicate time lags.

**Figure 5a**. Peak cross correlation time lag maps for 0-12 UT, 19 June 2010 for the field of view shown in Figure 2. The color bar on the bottom indicates the time lag range in seconds. The channel pair is indicated on the top of each panel

**Figure 5b.** Same as 5a. Note that the 211-193 and 171-131 pairs have different color bars.

**Figure 6a** Peak cross correlation time lag maps for 0-2 UT, 19 June 2010. The color bar on the bottom indicates the time lag range in seconds. The channel pair is indicated on the top of each panel.

**Figure 6b.** Same as 6a. Note the 211-193 and 171-131 pairs have different color bars.

**Figure 7a.** 335-211 time lag maps for all 2-hr time series. Temporal window in hours UT indicated on top. Top left panel is the same as 335-211 map in Figure 6a.

**Figure 7b** Same as 7a.

**Figure A.1a** Histogram of results shown in Figure 5. Time lag measurements for the 0-12 UT window, 19 June 2010 for the field of view shown in Figure 2. The channel pair is indicated on the top of each panel. The x axis is time lag and the y axis is the log number of pixels (note that the y axis is not the same between panels). Total number of positive, zero and negative time lag pixels are indicated in their respective x range.

**Figure A.1b** Same as A.1a.

**Figure A.2** Histograms of results shown in Figure 7. Time lag measurements for 335-211 for all 2-hour time windows. Temporal window in hours UT indicated on top. Total number of positive, zero and negative time lag pixels are indicated in their respective x range.

**Figure A.3a** Histogram of cross correlation value of the time lag results shown in Figures 5 and A.1. The channel pair is indicated on the top of each panel. Note that the y axis is not the same between panels.

**Figure A.3b** Same as A.3.a.

**Figure A.4** Histogram of cross correlation value of the time lag results shown in Figures 7 and A.2. Temporal window in hours UT indicated on top.

**Figure A.5** Top left panel: Map of the cross correlation value associated with the 94-335, 0-12 UT window, time lag map shown in Figure 5a. Top right panel: Time lag map as shown in Figure 5a; all pixels with cross correlation values less than or equal to 0.2 are blacked out. Lower panel: histogram of time lags at pixels where the cross correlation value is greater than 0.2.

Figure 1

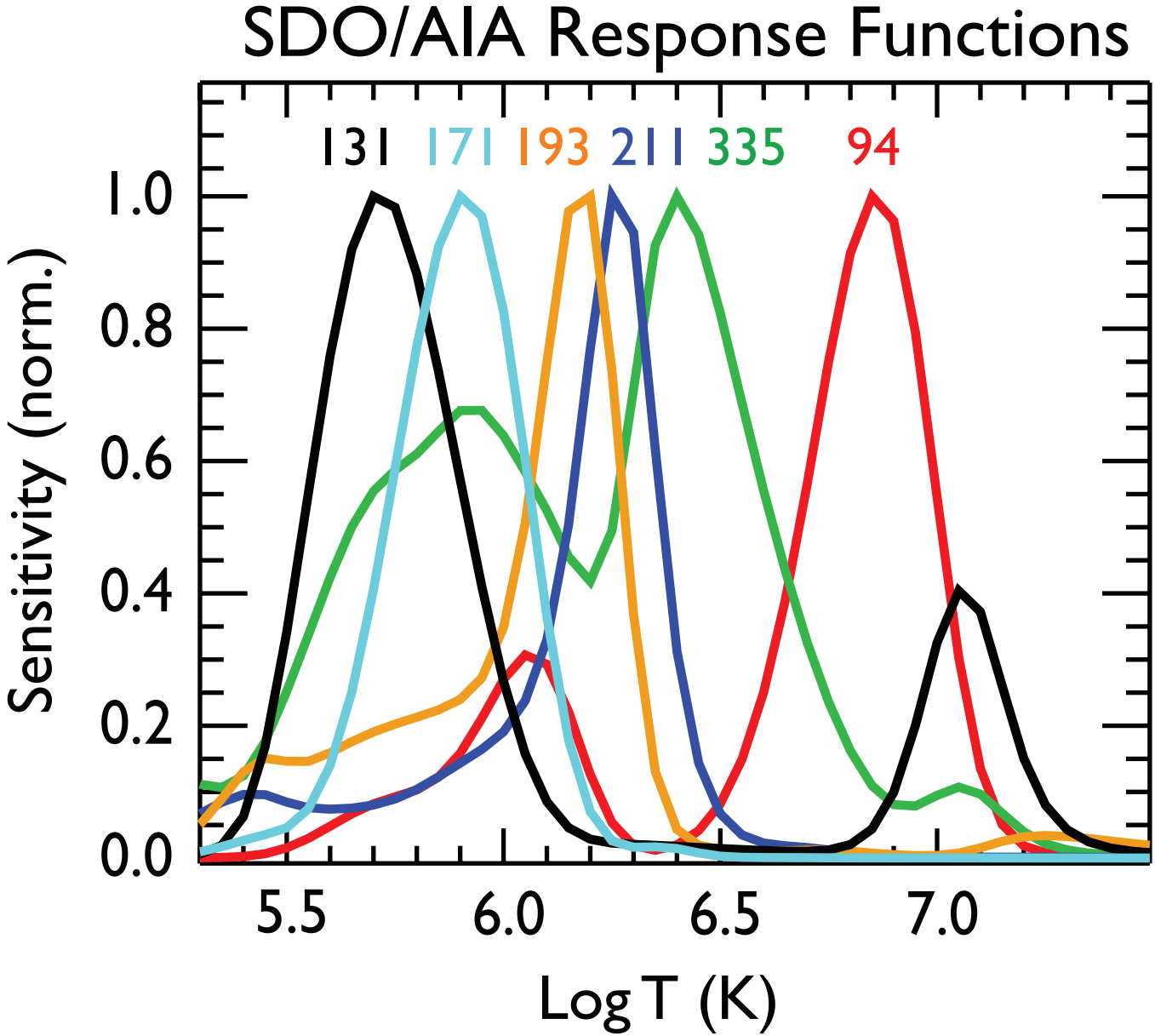

Figure 2

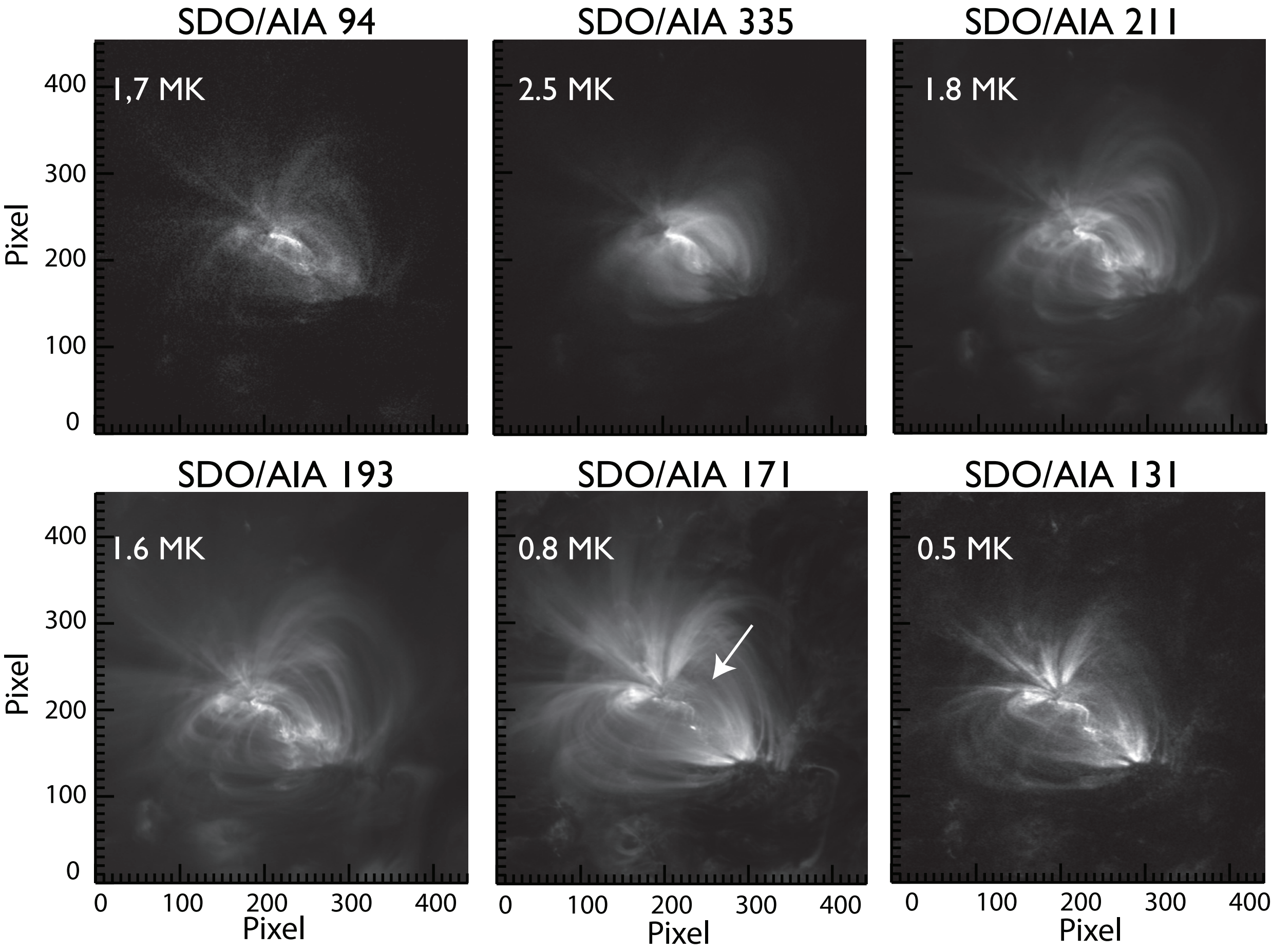

Figure 3

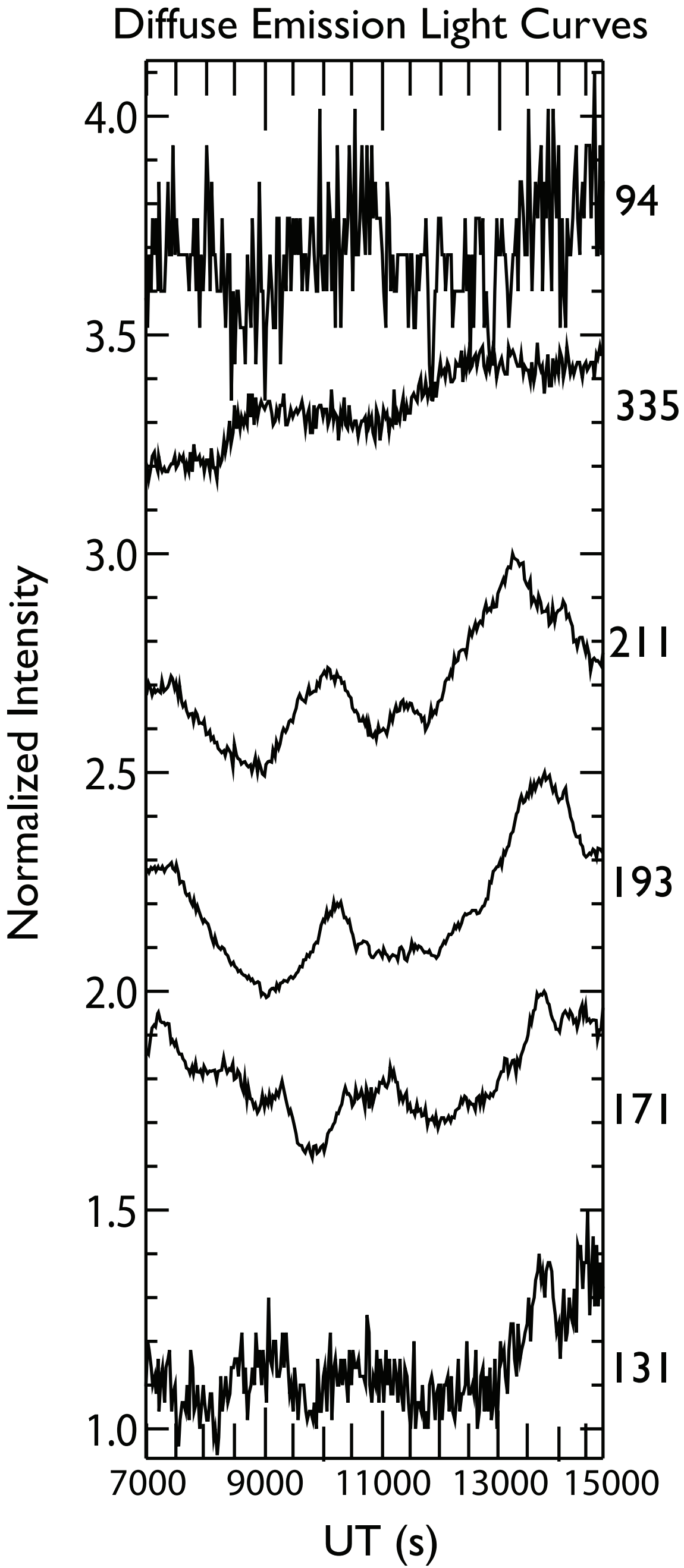

Figure 4

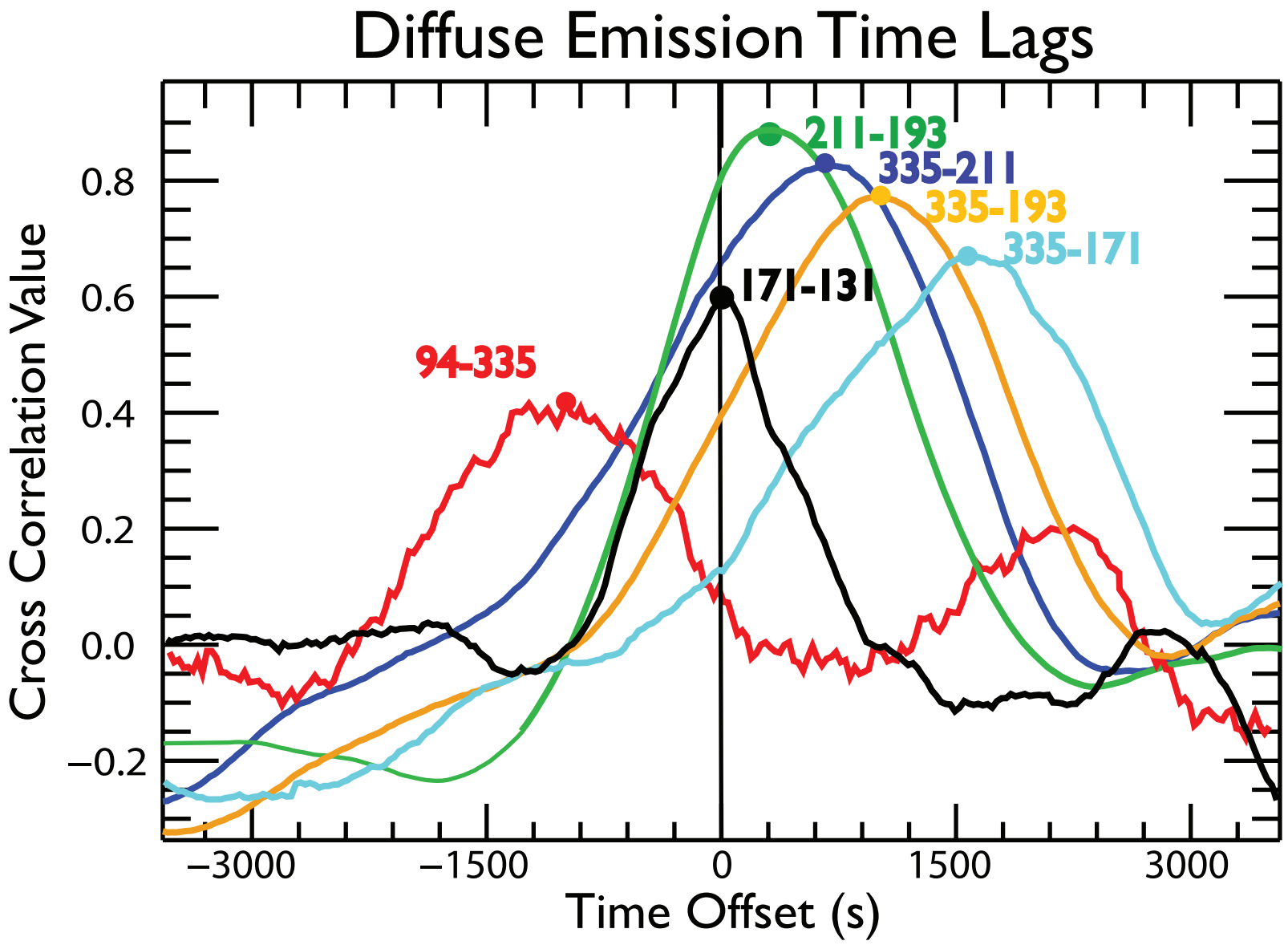

Figure 5a

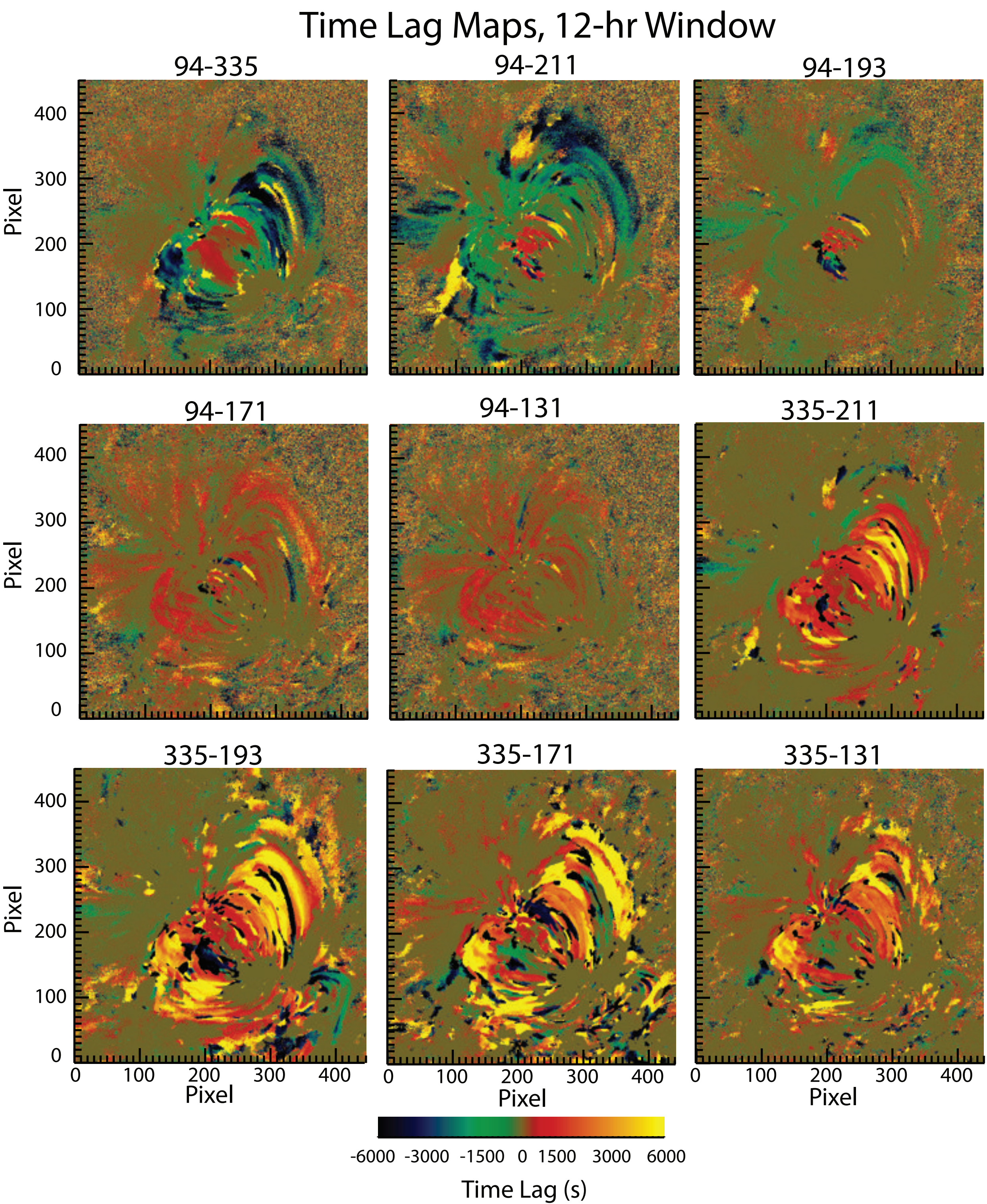

Figure 5b

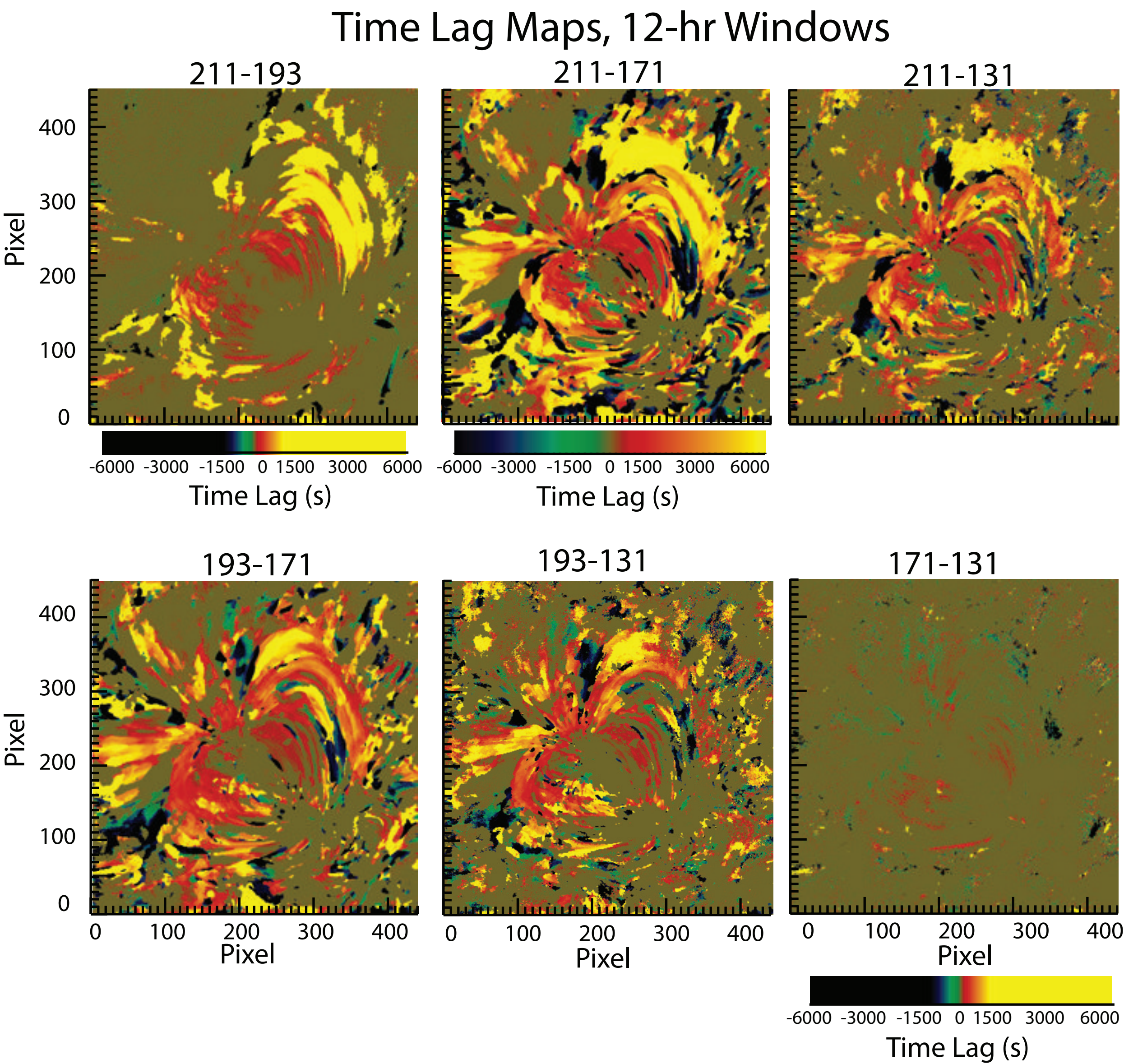

Figure 6a

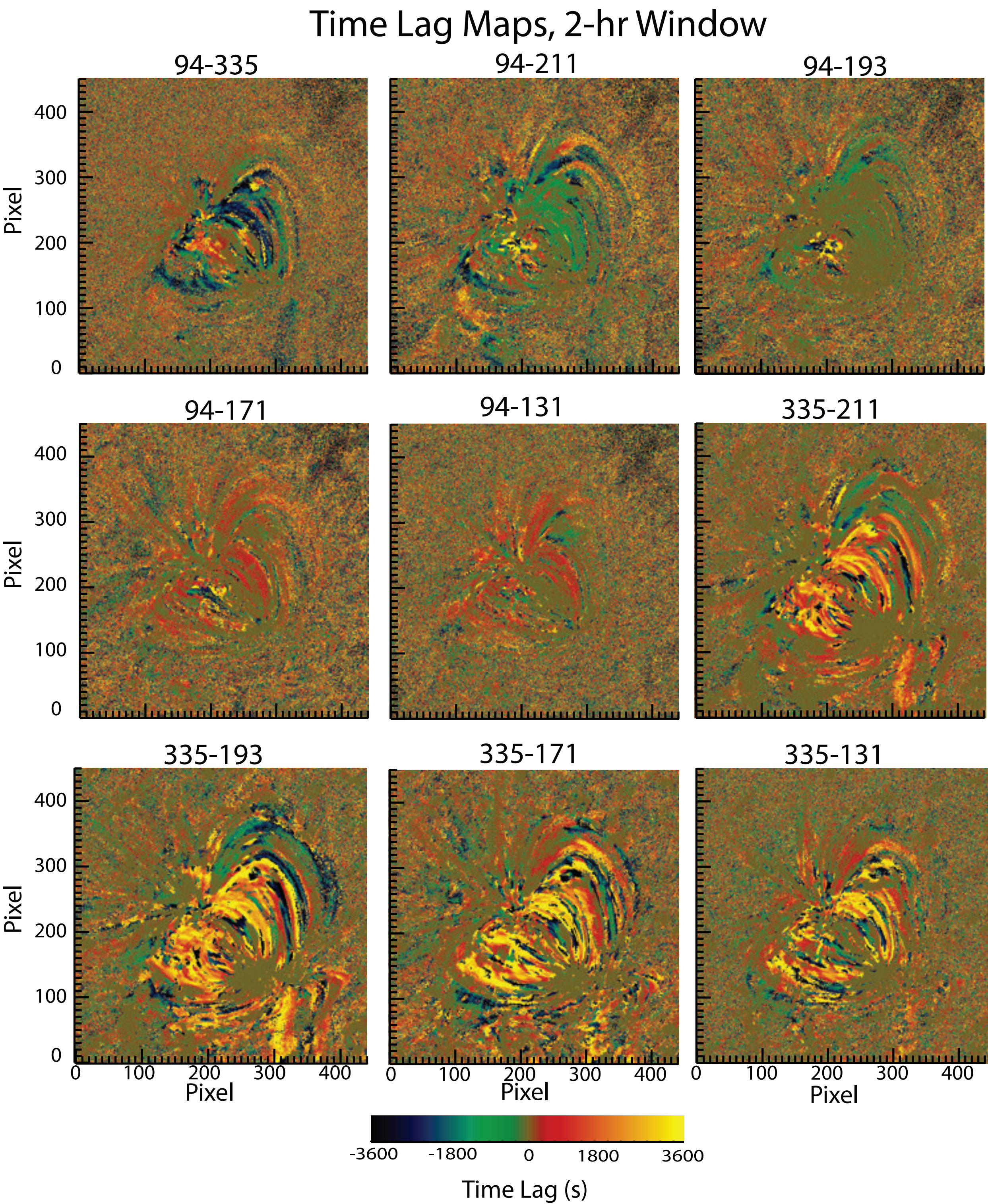

Figure 6b

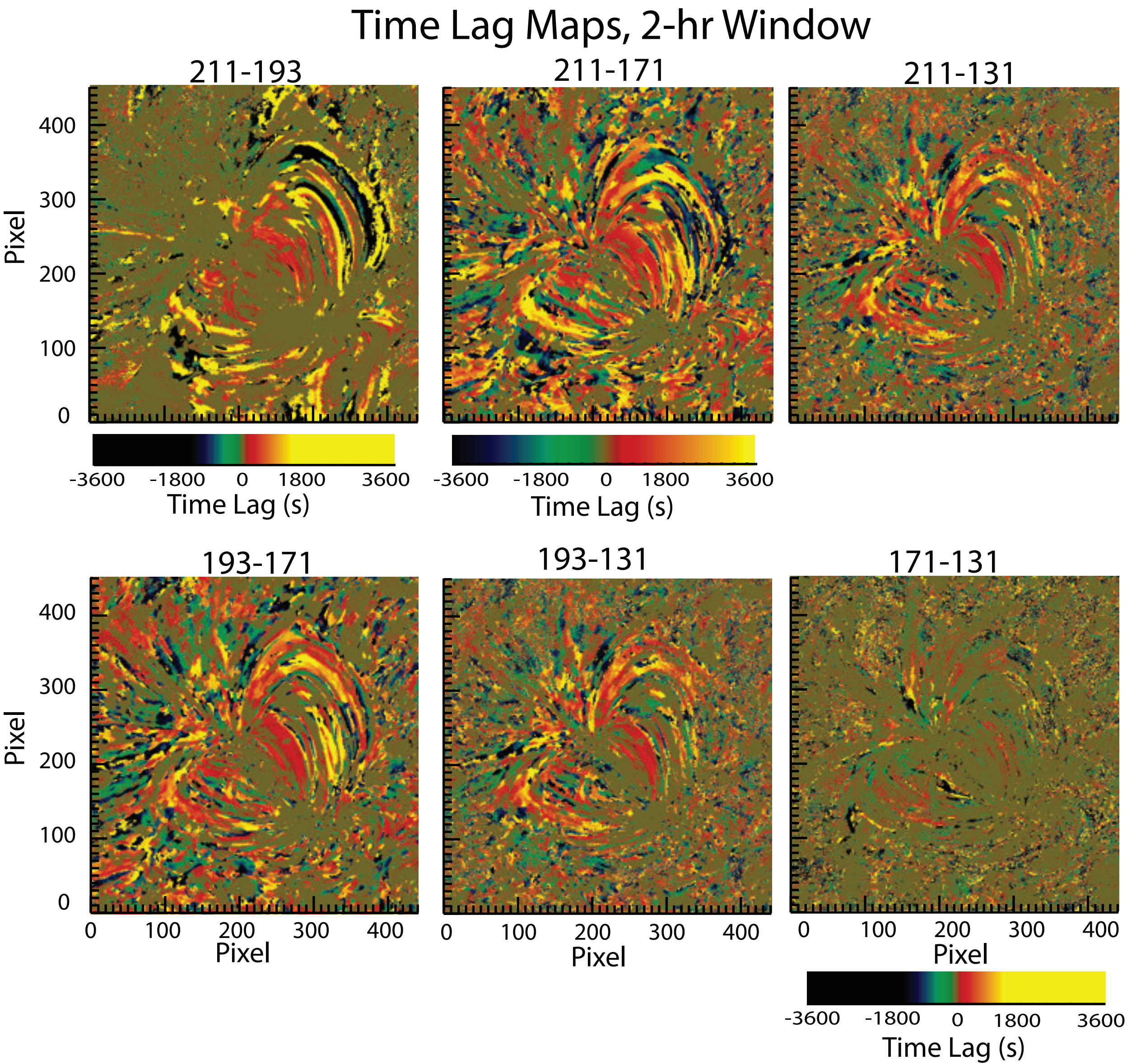

Figure 7a

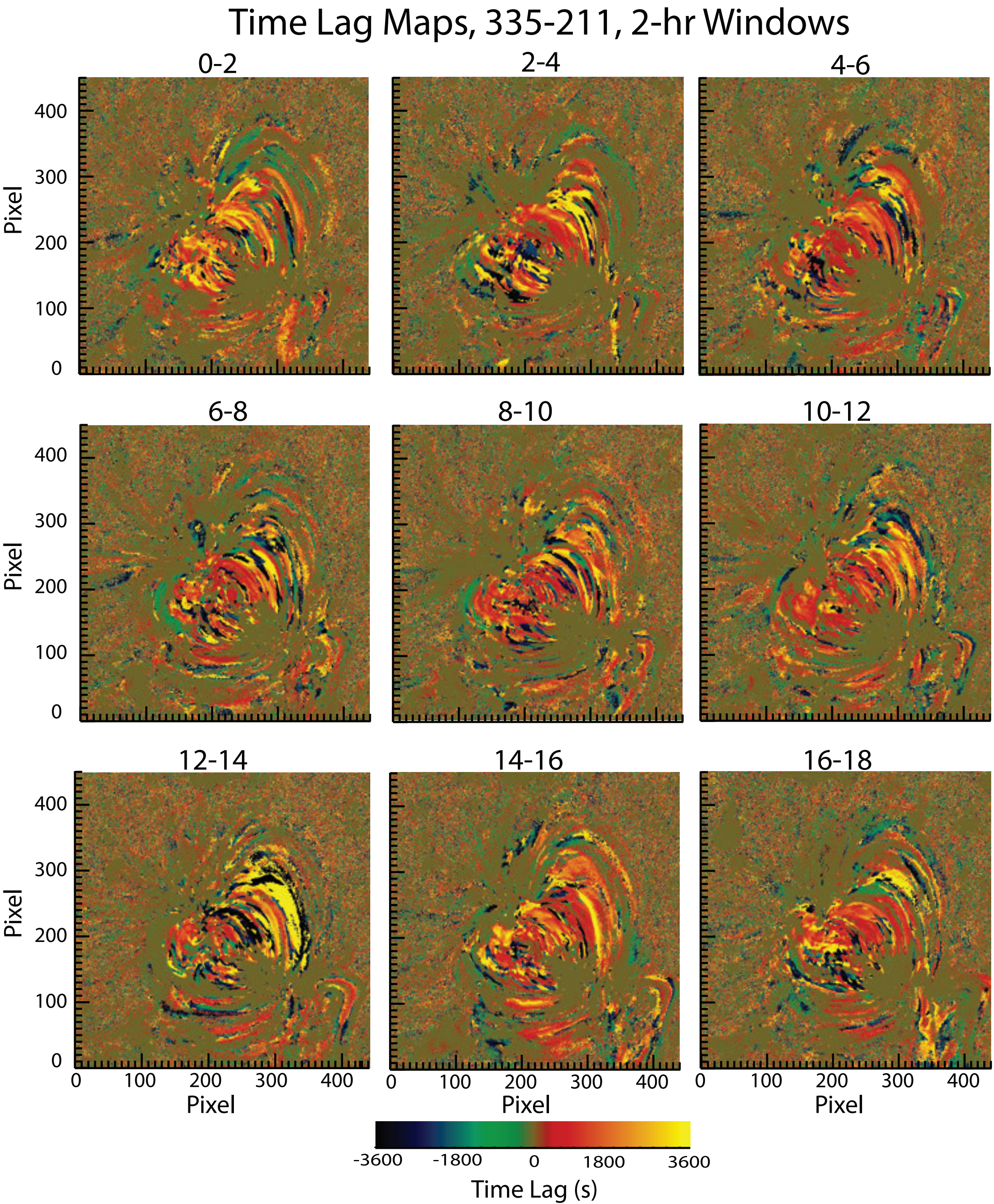

Figure 7b

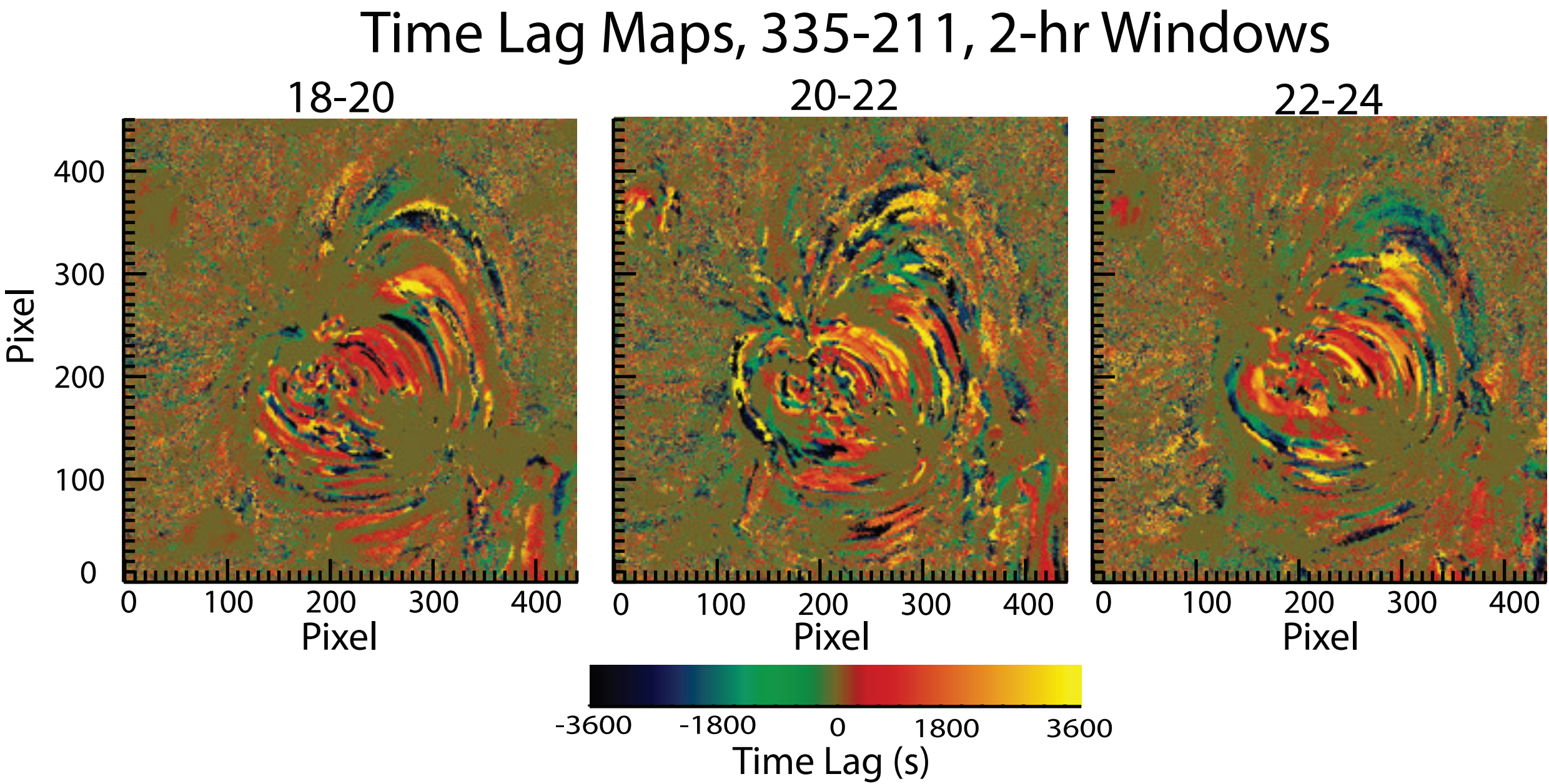

Figure A.1a

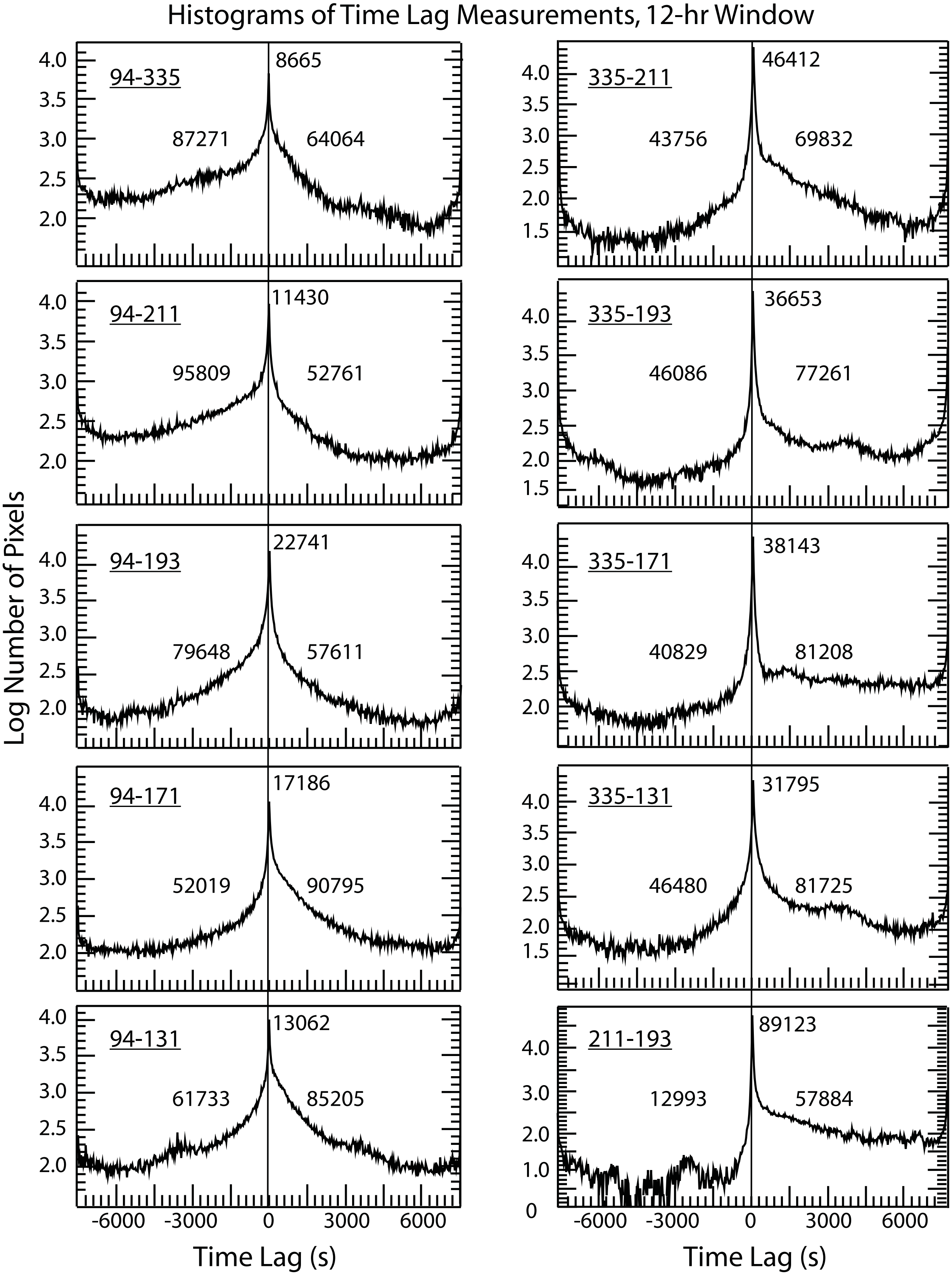

Figure A.1b

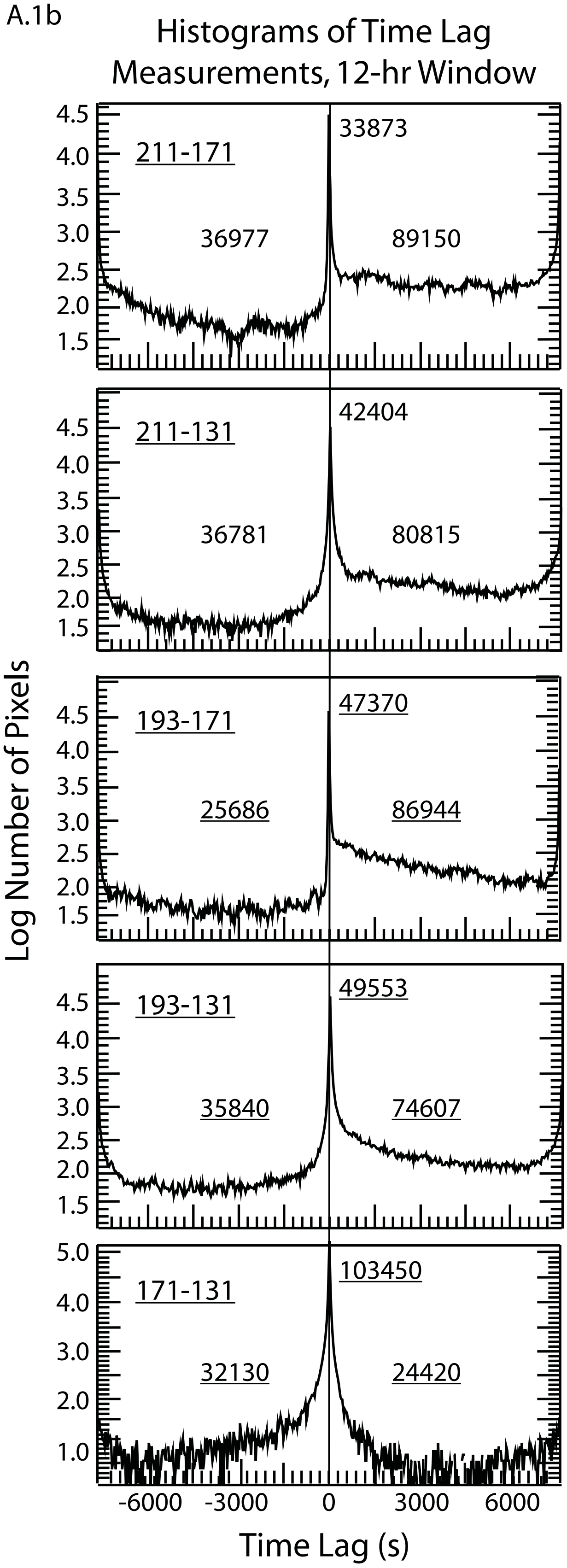

Figure A.2

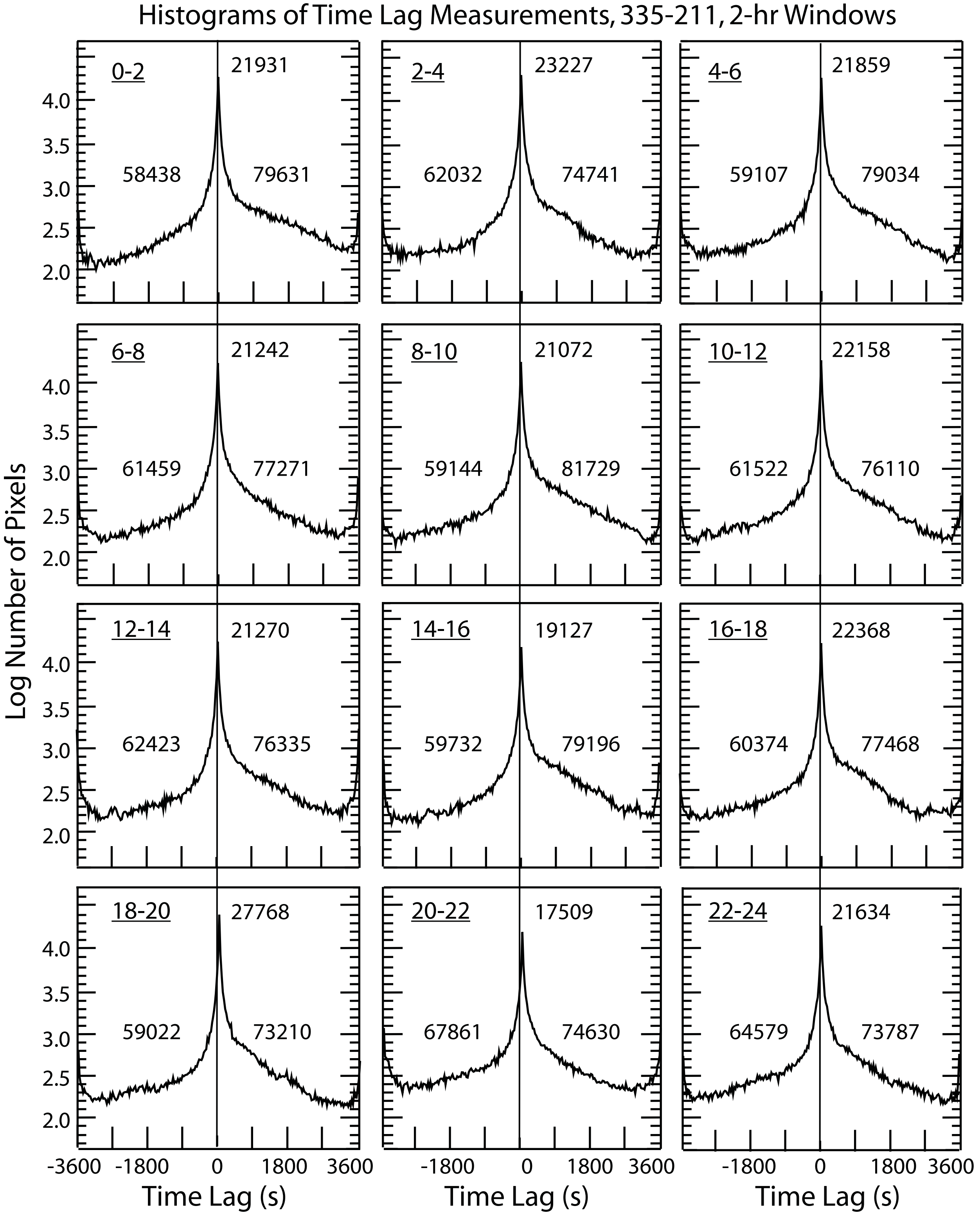

Histograms of Time Lag Measurements, 335-211, 2-hr Windows
Log Number of Pixels
4.0
3.5
3.0
2.5
2.0
0-2
21931
58438
79631
2-4
23227
62032
74741
4-6
21859
59107
79034
6-8
21242
61459
77271
8-10
21072
59144
81729
10-12
22158
61522
76110
12-14
21270
62423
76335
14-16
19127
59732
79196
16-18
22368
60374
77468
18-20
27768
59022
73210
20-22
17509
67861
74630
22-24
21634
64579
73787
-3600
-1800
0
1800
3600
Time Lag (s)

Figure A.3.a

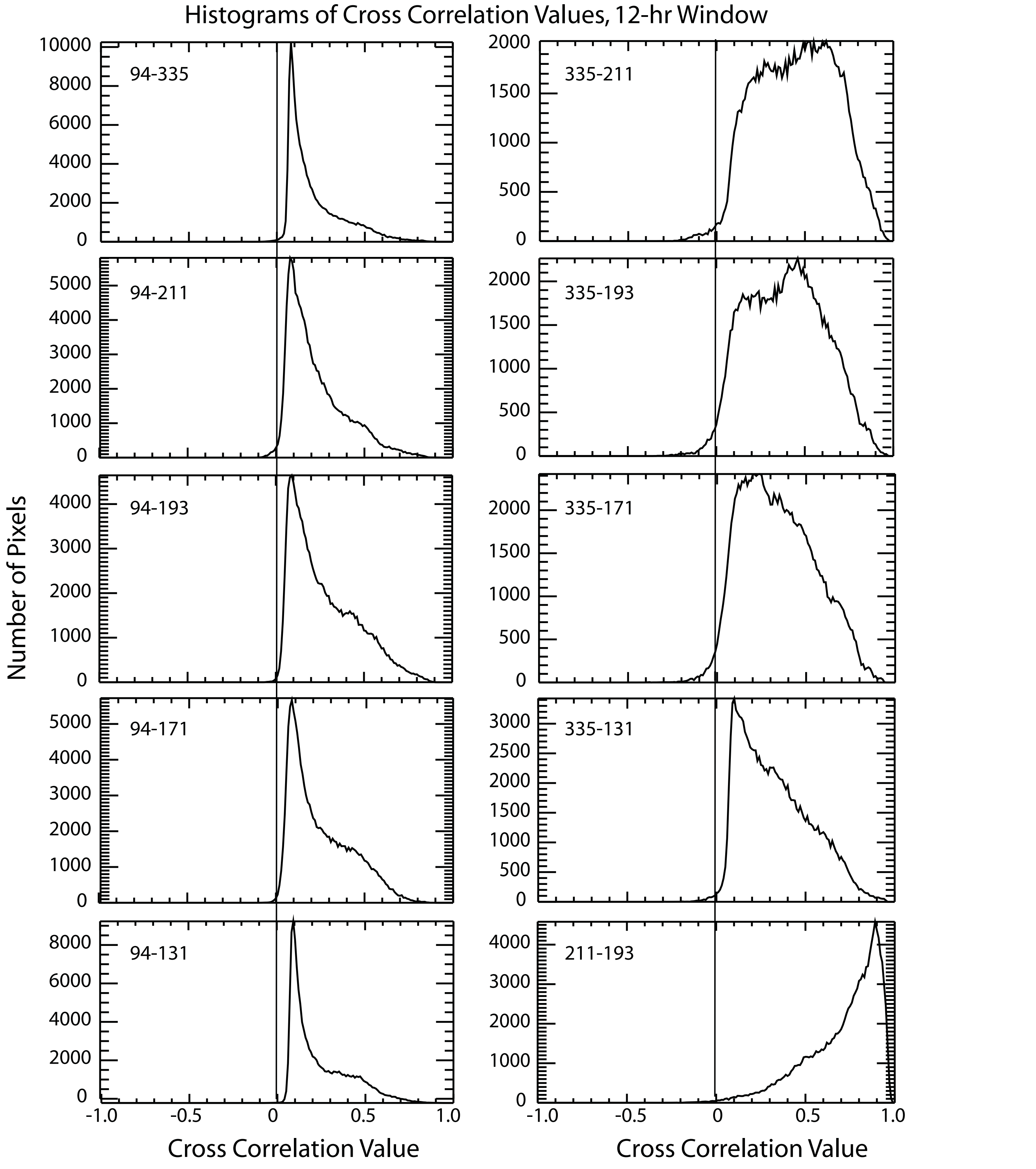

Figure A.3.b

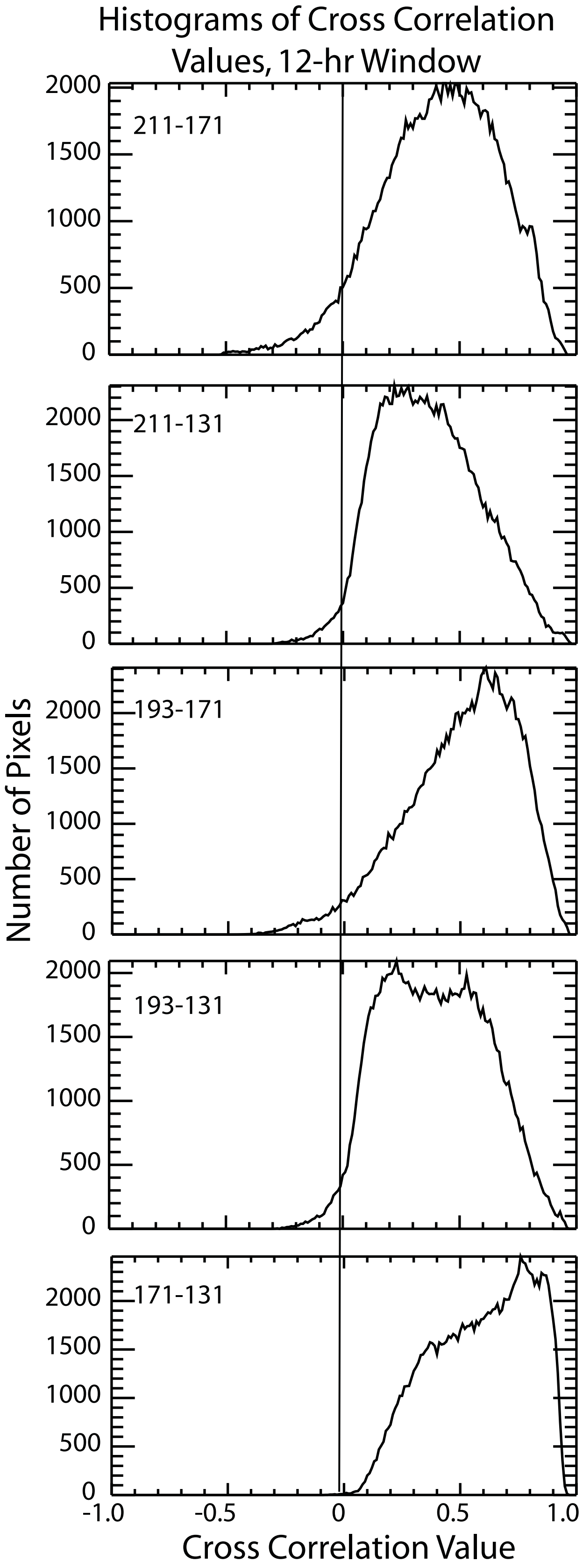

Figure A.4

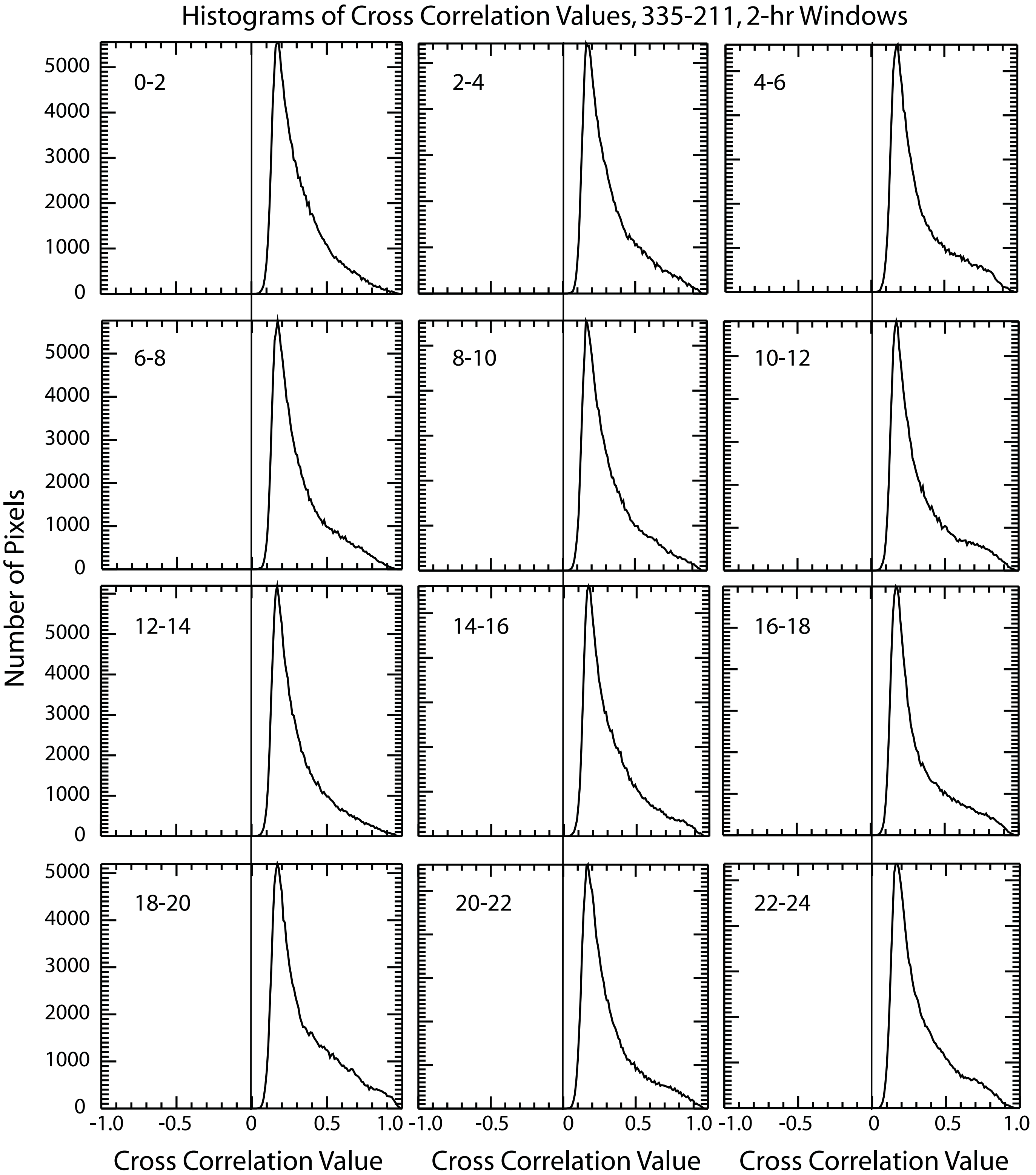

Histograms of Cross Correlation Values, 335-211, 2-hr Windows
Number of Pixels
Cross Correlation Value
0-2
2-4
4-6
6-8
8-10
10-12
12-14
14-16
16-18
18-20
20-22
22-24
0
1000
2000
3000
4000
5000
-1.0
-0.5
0
0.5
1.0

Figure A.5

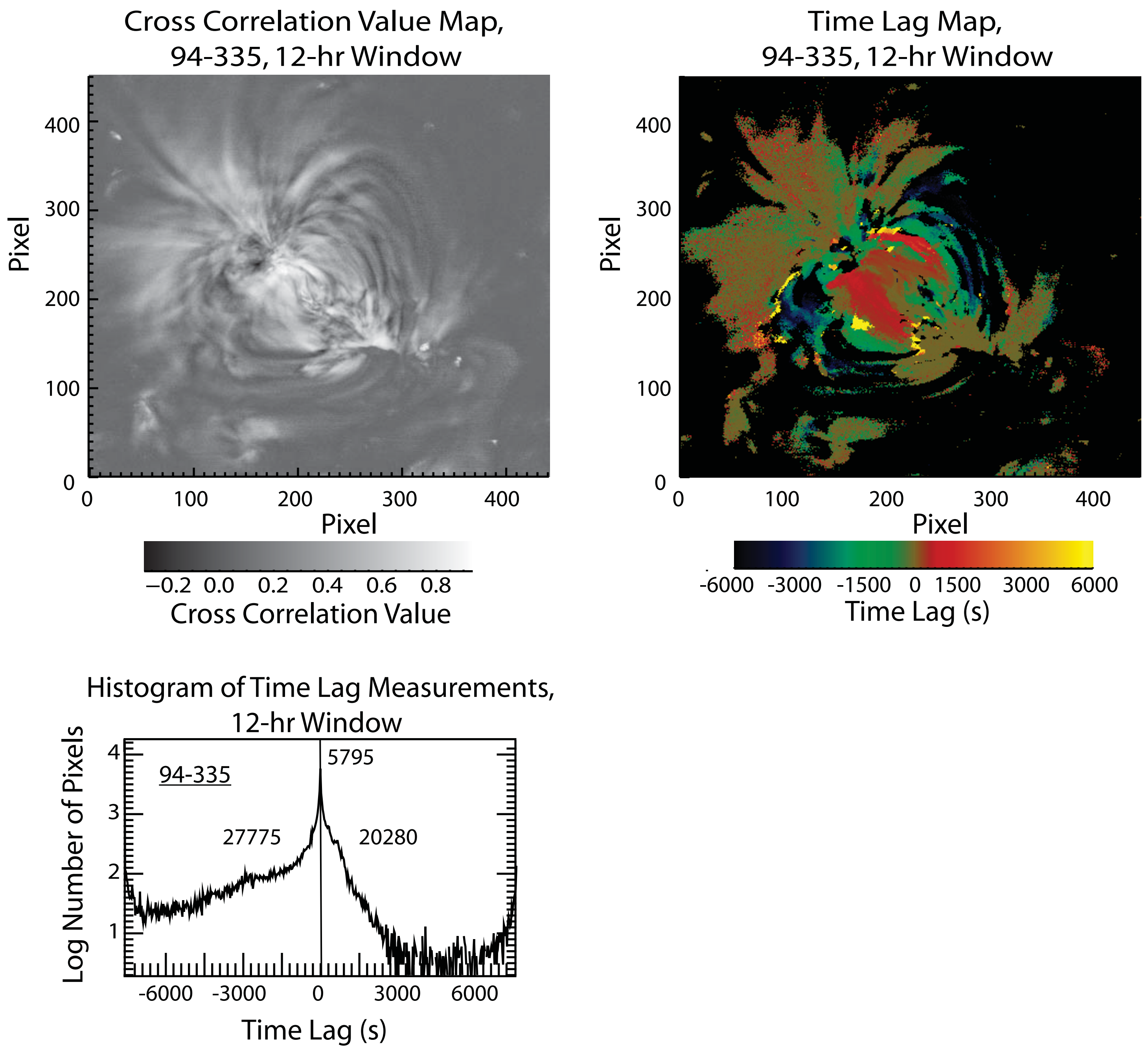